\documentclass[aps,superscriptaddress,showpacs,preprintnumbers,twocolumn]{revtex4} 
\usepackage{graphicx,epsfig,pstricks}
\usepackage{amssymb}
\usepackage{amsmath}
\usepackage{dsfont}
\usepackage{epstopdf}

\begin{document}

\title{Mott dissociation of pions and kaons in hot, dense quark matter}

\author{A.~Dubinin} \email{aleksandr.dubinin@ift.uni.wroc.pl}
\affiliation{Institute for Theoretical Physics, University of Wroc{\l}aw, 
50-204 Wroc{\l}aw, Poland} 

\author{A.~Radzhabov} \email{aradzh@icc.ru}
\affiliation{Matrosov Institute for System Dynamics 
and Control Theory, Irkutsk 664033, Russia}

\author{D.~Blaschke} \email{blaschke@ift.uni.wroc.pl}
\affiliation{Institute for Theoretical Physics, University of Wroc{\l}aw, 50-204 Wroc{\l}aw, Poland} 
\affiliation{Bogoliubov Laboratory for Theoretical Physics, JINR Dubna, 
141980 Dubna, Russia}
\affiliation{National Research Nuclear University (MEPhI), 115409 Moscow, Russia  }

\author{A.~Wergieluk} \email{a.wergieluk@gmail.com}
\affiliation{Department of Physics and Astronomy, University of California, 
Los Angeles, CA 90095-1547, USA}

\date{\today}

\begin{abstract}
We describe the Mott dissociation of pions and kaons within a Beth-Uhlenbeck approach based on the PNJL model, which allows for a unified description of bound, resonant and scattering states. 
Within this model we evaluate the temperature and chemical potential dependent modification of the phase shifts both in the pseudoscalar and scalar isovector meson channels for $N_f=2+1$ quark flavors. 
We show that the character change of the pseudoscalar bound states to resonances in the continuum at the Mott transition temperature is signaled by a jump of the phase shift at the threshold from $\pi$ to zero, in accordance with the Levinson theorem. 
In particular, we demonstrate the importance of accounting for the scattering continuum states, which ensures that the total phase shift in each of the meson channels vanishes at high energies, thus eliminating mesonic correlations from the thermodynamics at high temperatures. 
In this way, we prove that the present approach provides a unified description of the transition from a meson gas to a quark-gluon plasma.
We discuss the occurrence of an anomalous mode for mesons composed of quarks with unequal masses which is particularly pronounced for $K^+$ and $\kappa^+$ states at finite densities a a possible mechanism to explain the "horn" effect for the $K^+/\pi^+$ ratio in heavy-ion collisions. 
\end{abstract}
\maketitle

\section{Introduction}

Recently, the relativistic Beth-Uhlenbeck (BU) approach for two-particle
correlations in two-flavor quark matter within the Nambu--Jona-Lasinio (NJL) model 
\cite{Hufner:1994ma,Zhuang:1994dw,Blaschke:2013zaa,Dubinin:2013yga}
and the Polyakov-loop improved NJL (PNJL) model 
\cite{Wergieluk:2012gd,Yamazaki:2012ux,Yamazaki:2013yua,Blaschke:2014zsa,Dubinin:2015glr,Torres-Rincon:2016ahl} 
has appeared as a promising candidate for a unified description of hadron and quark-gluon matter.
In particular, the description of correlations in deconfined quark matter, sometimes denoted as 
"bound states above $T_c$" \cite{Ratti:2011au,Nahrgang:2013xaa,Nahrgang:2016lst}, 
with $T_c$ being the pseudocritical temperature of the chiral and 
deconfinement crossover transitions, is naturally achieved within this formulation. 
Attempts to formulate a unified thermodynamics of the hadron-to-quark-gluon matter transition have been made in postulating a spectral function for hadrons across the chiral/deconfinement transition 
\cite{Blaschke:2003ut,Turko:2011gw,Blaschke:2015nma}.

In order to reach this description, however, there are a few more important steps to be made. 
One of them is the inclusion of the strangeness degree of freedom.  
We devote the present work to this aim, restricting ourselves to the case of the lightest hadron
channels only, the pseudoscalar $\pi$ and $K$ mesons and their chiral partner states, 
the corresponding scalar mesons.

As is widely known, the NJL model is capable of describing the chiral restoration transition
in a hot and dense medium, where the dynamically generated quark masses drop as a function of temperature and chemical potentials, thus restoring the mass degeneracy of the chiral partner states.
At the same time, the continuum thresholds for quark-antiquark scattering channels drop, which results 
in a lowering of the binding energy for the pseudoscalar meson bound states and finally in their dissociation when they enter the continuum and change their character to resonances with a finite
lifetime (Mott effect). 
This aspect of the Mott dissociation is usually identified by mass poles for mesons becoming complex,
with the real part being the "mass" $M_i$ of the resonance $i$ and the imaginary part being related to a finite width $\Gamma_i$ due to its decay into the quark constituents. 
The phase shift of these states is well described by a Breit-Wigner function as long as $\Gamma_i\ll M_i$.    

Following previous works on the BU approach to pion dissociation in quark matter in the NJL 
\cite{Hufner:1994ma,Zhuang:1994dw,Blaschke:2013zaa,Dubinin:2013yga} and PNJL \cite{Wergieluk:2012gd,Yamazaki:2012ux,Yamazaki:2013yua,Blaschke:2014zsa,Dubinin:2015glr,Torres-Rincon:2016ahl} models,
we point out that the Breit-Wigner description is not complete as it neglects the fact that analytic properties of continuum states (resonances or just scattering states) are properly described by Jost functions, defined along the cut for continuum states in the complex energy plane. 
This fact is most appropriately taken into account by the introduction of a scattering phase shift function for each mesonic channel $\delta_i(s)$, which depends on the squared center-of-mass energy $s$ in the quark-antiquark system and thermodynamic parameters of the medium, here the temperature $T$ and chemical potentials $\mu_u=\mu_d=\mu$ and $\mu_s$ for light and strange quarks, respectively. 

The behaviour of the phase shift at the threshold can be used as an indicator for the Mott transition of a bound state to the scattering state continuum. 
The phase shift vanishes at infinity, while it has a value of $\pi$ at the continuum threshold as long as there is a bound state below the continuum.
When the bound state merges with the scattering states continuum, the phase shift jumps to zero, in accordance with the Levinson theorem.

A striking advantage of the BU approach over phenomenological models like \cite{Blaschke:2003ut,Turko:2011gw,Blaschke:2015nma} is the formulation of the thermodynamics of correlations in terms of phase shift functions that are in accordance with the  Levinson theorem \cite{Hufner:1994ma,Zhuang:1994dw,Wergieluk:2012gd,Blaschke:2013zaa,Dubinin:2013yga,Blaschke:2014zsa,Dubinin:2015glr},
which guarantees that the partial pressure of correlations vanishes at asymptotic temperatures (and chemical potentials). It may be understood as an anticipation of asymptotic freedom for strong, finite range interactions \cite{Dashen:1969ep}.

As a result, within the present approach the EoS and thermodynamic properties (like the composition) of the system of hadrons, quarks and gluons can be described with the correct asymptotics of the hadron gas (of pions and kaons) at low temperatures and the quark-gluon plasma at high temperatures, with the transition in-between. 

A quantitative comparison with lattice QCD thermodynamics data is premature at this stage of the development of the BU approach, mainly due to the lack of hadronic states and missing self-consistency.
It is therefore deferred to a future stage of work.

In this work we investigate thermodynamic properties of quark-gluon-meson matter for the PNJL model with $N_{f}=2+1$ quark flavors, where the strange chemical potential is fixed to $\mu_s=0.2~\mu$ as motivated by statistical model analyses of chemical 
freeze-out parameters for describing hadron production in heavy-ion collision experiments.
We study the temperature dependence of quark and meson masses along lines of constant $\mu/T$ in the $T$-$\mu$ plane which are seen as approximations to lines of constant entropy per baryon along which the hydrodynamic evolution of the heavy-ion collision shall proceed.
The EoS for pressure versus temperature in the Beth-Uhlenbeck approach is evaluated for 
$\mu/T=0,~0.5,~1.0,~2.0$, whereby the quark, gluon and meson contribution to the total pressure is given.
Special emphasis is on the Mott dissociation of the pseudoscalar meson states which is illustrated by the behaviour of the corresponding phase shifts as functions of energy for different temperatures and chemical potentials. 
The corresponding scalar meson states are unbound already in the vacuum, as is characteristic for the local PNJL model. 
Nevertheless one can observe how at high temperatures and chemical potentials the chiral partner states become degenerate in their analytic properties as encoded by the phase shifts. 

The paper is organized as follows. 
In Section \ref{sec:partition} we outline the path integral approach to the partition function of the 2+1 flavor PNJL model, including mesonic fluctuations beyond the meanfield in the scalar and pseudoscalar channels in Gaussian approximation.  
In Section \ref{sec:results} we present the results for the phase diagram and for the temperature dependence of the order parameters $\Phi$, $\bar{\Phi}$, $m_u=m_d$ and $m_s$ as well as the meson mass spectra, the energy dependent phase shifts and the pressure with its contributions from quarks, gluons and mesons along a set of trajectories of constant $\mu/T$ in the phase diagram.
In Section \ref{sec:conclusion} we present the conclusions of this study. 

\section{Partition function of the $2+1$ flavor PNJL model}
\label{sec:partition}

We employ a  $2+1$ flavor chiral quark model with NJL - type interaction kernel
based on the one used in works on the SU$_f$(3) scalar and 
pseudoscalar meson spectrum \cite{Costa:2002gk,Costa:2003uu,Costa:2005cz} 
developed on the basis of Ref.~\cite{Rehberg:1995kh} and its 
generalization by coupling to the Polyakov loop 
\cite{Hansen:2006ee,Costa:2008dp},
\begin{eqnarray}
\label{lagr}
{\cal L} &=& \bar{q} \left( i \gamma^\mu D_\mu + \hat{m}_0\right) q +
G_S \sum_{a=0}^{8} \left[ \left( \bar{q} \lambda^a q\right)^2+
 \left( \bar{q} i \gamma_5 \lambda^a q\right)^2
\right]
\nonumber\\
&-& \mathcal{U}\left(\Phi[A],\bar\Phi[A];T\right). 
\end{eqnarray}
Here $q$ denotes the quark field with three flavors, 
$f=u,d,s$, and three colors, $N_c=3$; $\lambda^a$ are the flavor SU$_f$(3) 
Gell-Mann matrices ($a=0,1,2,\ldots,8$), $G_S$ is a coupling constant.
The scalar-pseudoscalar meson interaction channels in (1) are color
singlet and fulfill the requirement of being chirally symmetric. 
Thus, only the diagonal matrix of current quark masses 
$\hat{m}_0 = \mbox{diag}(m_{0,u}, m_{0,d}, m_{0,s})$ induces an explicit breaking
of the otherwise global symmetry of the Lagrangian (\ref{lagr}).
This is a property shared with the QCD Lagrangian.
The covariant derivative is defined as
$D^{\mu}=\partial^\mu-i A^\mu$, with $A^\mu=\delta^{\mu}_{0}A^0$ 
(Polyakov gauge); in Euclidean notation $A^0 = -iA_4$.  
The strong coupling constant $g_s$ is absorbed in the definition of 
$A^\mu(x) = g_s {\cal A}^\mu_a(x)\frac{\lambda_a}{2}$, where 
${\cal A}^\mu_a$ is the (SU$_c$(3)) gauge field and $\lambda_a$ are the 
Gell-Mann matrices in SU$_c$(3) color space.

The Polyakov loop field  $\Phi$ appearing in the potential term of
(\ref{lagr}) is related to the gauge field through the gauge covariant
average of the Polyakov line~\cite{Ratti:2005jh}
\begin{equation}
\Phi(\vec x)=\left\langle \left\langle l(\vec x)\right\rangle\right\rangle
=\frac{1}{N_c}{\rm Tr}_c\left\langle \left\langle L(\vec x)
\right\rangle\right\rangle,
\label{eq:phi}
\end{equation}
where
\begin{equation}
L(\vec x) ={\cal P}\exp\left[i\int_0^\beta d\tau A_4(\vec x, \tau)\right]\,.
\label{eq:loop}
\end{equation}
Concerning the effective potential for the (complex) $\Phi$ field, we adopt 
the polynomial form and the parametrization proposed in Ref.~\cite{Ratti:2005jh}.
Alternatively, the logarithmic form of this potential \cite{Roessner:2006xn}
could also be used. 

The (P)NJL model is a primer for describing the dynamical breakdown of the
(approximate) chiral symmetry in the vacuum and its partial restoration at high
temperatures and chemical potentials. 
This feature is governed by the occurrence of a nonvanishing expectation value 
for the mean field in the scalar meson channel. 
At the same time this model provides a field-theoretic description of 
pseudoscalar meson properties which is in accordance with the low energy 
theorems of QCD, such as the Goldstone theorem.  
This is achieved by considering the (Gaussian) fluctuations above the mean field
in the scalar and pseudoscalar meson sector and analysing their properties as
encoded in the matrix elements of the  polarization operator which after analytic 
continuation from the bosonic Matsubara frequencies to the real $q_0$ axis reads
\cite{Klevansky:1992qe}

\begin{eqnarray}
\Pi^{M^a}_{ff'} (q_0 ,{\bf q}) &=& 2N_cT \sum_{n} \int \frac{d^3p}{(2\pi)^3}\mbox{tr}_{D}
\left[ S_{f}(p_n ,{\bf p}) \Gamma^{M^a}_{ff'} \right.\nonumber\\
&&\times \left.S_{f'}(p_n + q_0,{\bf p+q}) \Gamma^{M^a}_{ff'} \right],
\end{eqnarray} 
where
\begin{eqnarray}
\Gamma^{P^a}_{ff'}&=& i\gamma_5\ T^a_{ff'} ~,~~\Gamma^{S^a}_{ff'} = T^a_{ff'}~,\\
T^a_{ff'} &=&
\left\{
\begin{array}{l}
(\lambda_3)_{ff'}, \\
(\lambda_1\pm i \lambda_2)_{ff'}/\sqrt{2}, \\
(\lambda_4\pm i \lambda_5)_{ff'}/\sqrt{2}, \\
(\lambda_6\pm i \lambda_7)_{ff'}/\sqrt{2},
\end{array}
\right.
\end{eqnarray}
for $P^a=\pi^0, \pi^\pm, K^\pm, K^0, \bar{K}^0$ denoting the pseudoscalar meson states and 
$S^a=a_0^0, a_0^\pm, \kappa^\pm, \kappa^0, \bar{\kappa}^0$ the scalar meson states.
Here $\mbox{tr}_{D}$ is the trace over Dirac matrices, 
the sum over $n$ denotes the sum over the fermionic Matsubara frequencies $\omega_n=(2n+1)\pi T$,  
and
$S_f (p_n,{\bf p})=[\gamma_0(i\omega_n+\mu_f + A^0)-{\bf \gamma}{\bf p} - m_f]^{-1}$ is the quark 
Green function with the dynamical quark mass $m_f$ and chemical potential $\mu_f$ for the flavor $f$. 

The matrix elements of the polarization operator can be represented in terms of two integrals which after summation over the Matsubara frequencies for mesons at rest in the medium (${\bf q}={\bf 0}$) are given by 
\begin{eqnarray}
&&\Pi^{P^a,S^a}_{ff'} (q_0+i\eta, {\bf 0}) = 4 \bigl\{  I_{1}^f(T,\mu_{f})+I_{1}^{f'}(T,\mu_{f'})  \nonumber  \\
&&\, \mp \left[  (q_0+\mu_{ff'})^2 -(m_f \mp m_{f'})^2 \right] I_2^{ff'}(z,T,\mu_{ff'}) \bigr\}\, , \nonumber
\end{eqnarray}
where $\mu_{ff'}=\mu_{f}-\mu_{f'}$.
To parametrize the model with known pseudoscalar meson 
masses the vacuum expressions of the integrals are used
\begin{eqnarray}
\label{i1}
I_1^f(0,0)
&=&\frac{N_c}{4 \pi^2} \int^{\Lambda}_0 \frac{dp \, p^2}{E_f} ,
\end{eqnarray}
{\color{black}
\begin{eqnarray}\label{i2}
I_2^{ff'}(z,0,0) 
       &=& \frac{N_c}{4 \pi^2} \int^{\Lambda}_0 \frac{dp \, p^2}{E_f E_{f'}}
       \frac{E_f+E_{f'}}{z^2-(E_f+ E_{f'})^2} ,
\end{eqnarray}}

with $E_{f}=\sqrt{{p}^2+m_{f}^2}$ being the quark energy.

The integrals for finite temperature and chemical potential have the following form
\begin{eqnarray}\label{firstt}
I_1^f(T,\mu_f) &=& \frac{N_c}{4\pi^2} \int_0^\Lambda \frac{dp \, p^2}{E_f} \left(n^-_f - n^+_f \right),
\\
I_2^{ff'} (z,T,\mu_{ff'}) &=& \frac{N_c}{4\pi^2} \int_0^\Lambda \frac{dp \, p^2}{E_fE_{f'}}
\nonumber\\
&&\Biggl[ \frac{E_{f'}}{(z-E_f-\mu_{ff'})^2-E_{f'}^2} \,\, n^-_f \nonumber \\
&&  -  \frac{E_{f'}}{(z+E_f-\mu_{ff'})^2-E_{f'}^2} \,\, n^+_f
\nonumber \\ 
&& +\frac{E_f}{(z+E_{f'}-\mu_{ff'})^2-E_f^2} \,\,  n^-_{f'} 
\nonumber \\
  && 
- \frac{E_f}{(z-E_{f'}-\mu_{ff'})^2-E_f^2} \,\,  n^+_{f'} 
\Biggr], 
\end{eqnarray}
where  $n_f^{\pm}=f^+_\Phi(\pm E_f)$ are the generalized fermion distribution 
functions \cite{Costa:2008dp, Blaschke:2014zsa} for quarks of flavor $f$ with positive (negative) 
energies in the presence of the Polyakov loop values $\Phi$ and $\bar{\Phi}$ 
\begin{eqnarray}
\label{f-Phi}
f^+_\Phi(E_f)&=&
\frac{(\bar{\Phi}+2{\Phi}Y)Y+Y^3}{1+3(\bar{\Phi}+{\Phi}Y)Y+Y^3}
~,\\
f^-_\Phi(E_f)&=&
\frac{({\Phi}+2\bar{\Phi}\bar{Y})\bar{Y}+\bar{Y}^3}{1+3({\Phi}+\bar{\Phi}\bar{Y})\bar{Y}+\bar{Y}^3}
~,
\label{f-Phi-bar}
\end{eqnarray}
where the abbreviations $Y={\rm e}^{-(E_f-\mu_f)/T}$ and $\bar{Y}={\rm e}^{-(E_f+\mu_f)/T}$ are used. 
The functions (\ref{f-Phi}) and (\ref{f-Phi-bar}) fulfill the relationship 
$f^+_\Phi(-E_f)=1-f^-_\Phi(E_f)$, 
and they go over to the ordinary Fermi functions $f_1^\pm(E_f)$ for $\Phi=\bar \Phi=1$, where
\begin{eqnarray}\label{fermi}
         f_1^\pm(E_f) = \frac{1}{1+ {\rm e}^{\beta (E_f \mp \mu_f)}}~.
\end{eqnarray} 

The quark masses $m_f$ are found by solving the gap equation
\begin{eqnarray}
m_f = m_{0,f} + 16\, m_f G_S I_1^f(T,\mu),
\end{eqnarray}
while the meson mass spectrum is obtained from the pole condition
\begin{eqnarray}\label{mass}
\mathcal{P}^{M^a}_{ff'}(M_{M^a}+i\eta, {\bf 0})=
1- 2G_S  \Pi^{M^a}_{ff'}(M_{M^a}+i\eta, {\bf 0}) =0.
\nonumber\\
\end{eqnarray}

Note that when there is no bound state solution below the continuum threshold for
$q_0=M_{M^a}<m_{{\rm thr}, ff'}$,  $m_{{\rm thr}, ff'} = m_f+m_{f'}$, 
then in the vicinity of the threshold, for $M_{M^a}>m_{{\rm thr},ff'}$, Eq.~(\ref{mass}) can still be solved in its 
complex form in order to determine the mass of the resonance $M_{M^a}$ and the 
respective decay width $\Gamma_{M^a}$. 
Thus, we assume that this equation can be written as a system of two coupled equations
(upper (lower) sign for $M^a=P^a\, (S^a)$)
\begin{eqnarray}
\label{BSE1}
&&M_{M^a}^2-\frac{1}{4}\Gamma_{M^a}^2-(m_f\mp m_{f'})^2 = -\mbox{Re} I_2(M_{M^a},T,\mu_{ff'})\nonumber\\
&&\hspace{1cm}\times \frac{(8G_S)^{-1}\mp(I_1^f(T,\mu_f)+I_1^{f'}(T,\mu_{f'}))}{|I_2(M_{M^a}+i\eta,T,\mu_{ff'})|^2} ~,
\label{BSE2}
\end{eqnarray}
\begin{eqnarray}
\label{BSE2}
M_{M^a} \Gamma_{M^a} &=& - \mbox{Im} I_2(M_{M^a}+i\eta,T,\mu_{ff'})
\nonumber \\  
 && \times \frac{(8G_S)^{-1}\mp (I_1^f(T,\mu_f)+I_1^{f'}(T,\mu_{f'}))}{|I_2(M_{M^a}+i\eta,T,\mu_{ff'})|^2}~,
 \nonumber\\
\end{eqnarray}
which have solutions of the form 
\begin{eqnarray}
q_0=M_{M^a}-i\frac{1}{2}\Gamma_{M^a}~. 
\end{eqnarray}

To study the thermodynamics we use the following expression for the quark pressure of flavor $f$  
\begin{eqnarray}
\label{pressure-quark}
P_f &=&
-\frac{(m_f-m_{0,f})^2}{8G}
+\frac{N_c}{\pi^2}\int_0^\Lambda {dp\, p^2}\, E_f
\nonumber\\
&+&\frac{N_c}{3\pi^2}\int_0^\infty\frac{dp\, p^4}{E_f}
	\left[f_\Phi^+(E_f) + f_\Phi^-(E_f)\right]~,
\end{eqnarray}
and for  the mesonic pressure we use the BU form in no-sea approximation with the phase shifts 
\begin{eqnarray}
\label{GBU_M}
P_{M}
	&=&
	d_{\rm M} \int\frac{{\rm d}^3q}{(2\pi)^3}~
	\int_0^\infty\frac{{\rm d}\omega}{2 \pi}~
\bigg\{\nonumber\\
&& g(\omega-\mu_{\rm M})+g(\omega+\mu_{\rm M})
	\bigg\}
	\delta_{\rm M}(\omega, {\bf q})
	~,
	\label{BUU}
\end{eqnarray}
where $\delta_{\rm M}(\omega, {\bf q})$ is the mesonic phase shift which has the form
\begin{eqnarray}
\delta_{\rm M}(\omega,{\bf q}) = -\arctan \left\{
\frac
{\mathrm{Im}\left(\mathcal{P}^M_{ff'}(\omega-i\eta,{\bf q})\right)}
{\mathrm{Re}\left(\mathcal{P}^M_{ff'}(\omega+i\eta,{\bf q})\right)}
\right\}~,
\end{eqnarray}
and $g(E)$ is the Bose function
\begin{eqnarray}
\label{Bose}
g(E)=\frac{1}{{\rm e}^{\beta E}-1}~.
\end{eqnarray}

Note that the physical mesons are color singlet states and therefore their distribution functions do not 
depend on the Polyakov loop.
This is in striking difference to the case of diquarks which are colored objects and their distribution is 
strongly suppressed by the coupling to the Polyakov loop in the confined domain.
In the deconfined domain, diquarks are suppressed too by the Mott dissociation to their quark constituents
\cite{Blaschke:2014zsa}. 
Therefore we do not include them in our considerations in this work.

\section{Results and Discussion}
\label{sec:results}

\begin{figure}[!t]
\centerline{\includegraphics[width=0.5\textwidth]{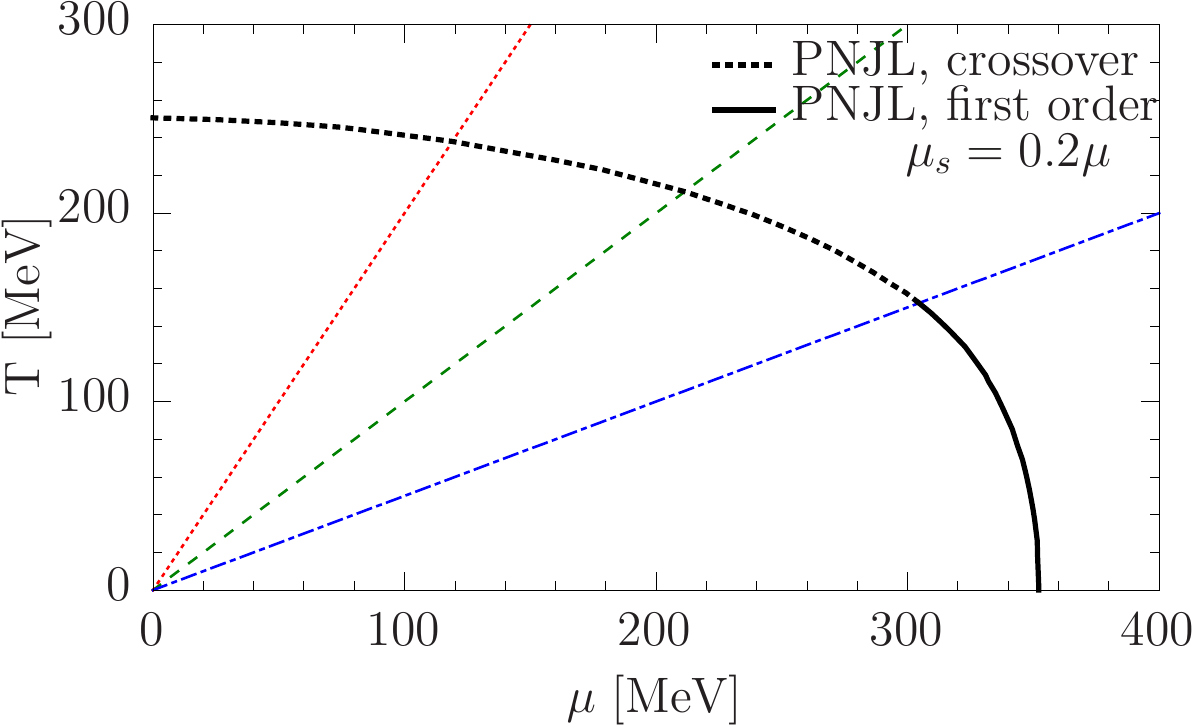}}%
\caption{Phase diagram of the PNJL model: the black solid line corresponds to the first order phase transition which turns into a crossover transition shown by the black dotted line. 
Three thin lines correspond to fixed ratios $\mu/T$= 0.5 (red dotted line), $\mu/T=1.0$ (green dashed line) and $\mu/T=2.0$ (blue dash-dotted line). 
}%
\label{Fig:Phasdig}%
\end{figure}

The parameters used for the numerical studies in this work are  the bare quark masses  
$m_{0(u,d)}= 5.5~$MeV and  $m_{0s}= 138.6~$MeV, the three-momentum cut-off  
$\Lambda= 602~$MeV and the scalar coupling constant $G_{S}\Lambda^{2}= 2.317$. 
With these parameters one finds in vacuum a constituent  quark mass for light quarks of 367 MeV, a pion mass of 135 MeV and pion decay constant $f_{\pi}=92.4~$MeV.  
We present the results along trajectories in the PNJL phase diagram shown in Fig.~\ref{Fig:Phasdig}.   Each trajectory corresponds to a constant ratio  $\mu/T = 0, 0.5, 1.0, 2.0$, where $\mu=\mu_u=\mu_d$ is the light quark chemical potential. 
For the strange quark chemical potential the relation $\mu_s=0.2~\mu$  is used \cite{Naskret:2015pna},
see also \cite{Karsch:2010ck}.     

\begin{figure}[!bh]
\centerline{\includegraphics[width=0.5\textwidth]{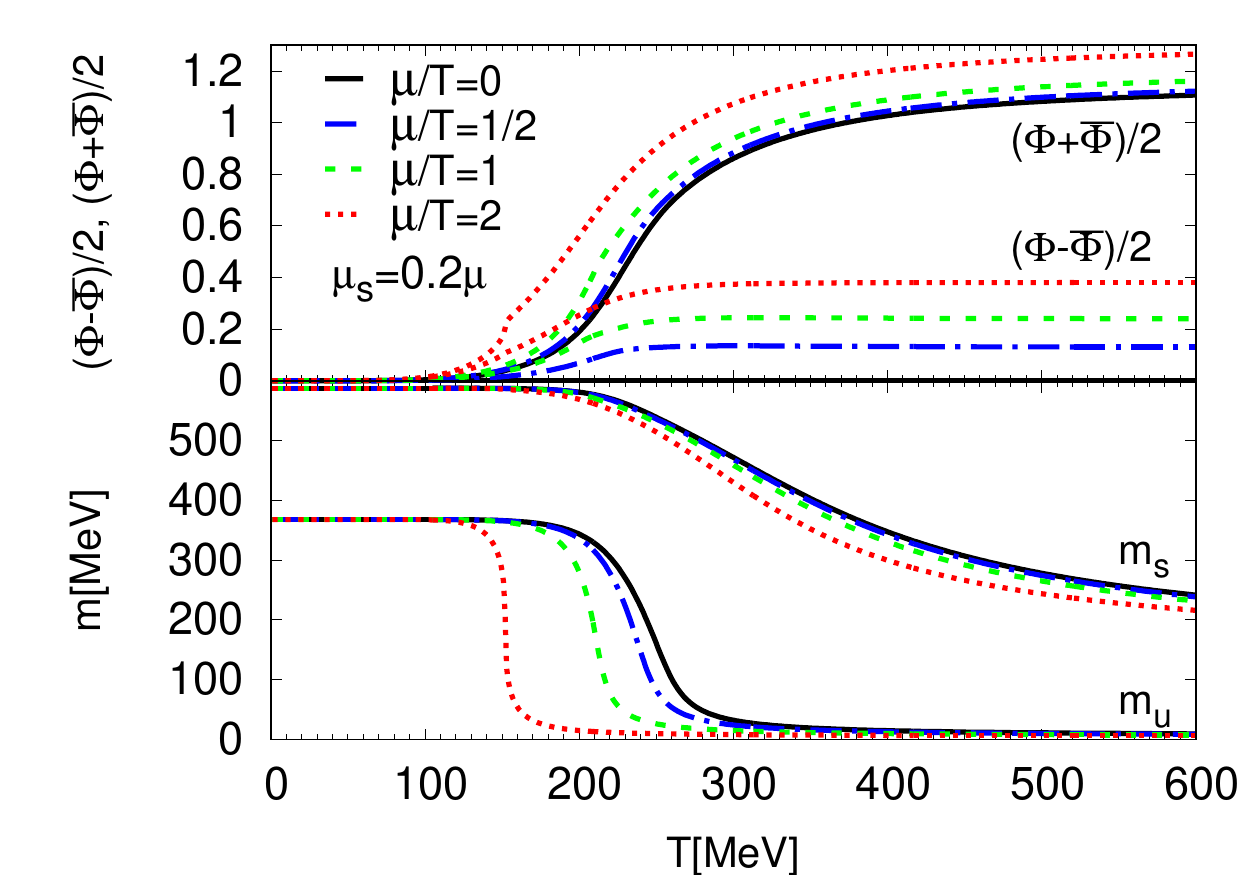}}%
\caption{Temperature dependence of the dynamical masses for light quarks $m_u=m_d$ and for the strange quark $m_s$ (lower panel) together with the sum and the difference of the traced Polyakov loop 
$\Phi$ and its conjugate $\bar{\Phi}$ (upper panel) along different trajectories in phase diagram: 
$\mu/T=0$ (solid black line), $\mu/T=0.5$ (blue dash-dotted line), 
$\mu/T=1.0$ (green dashed line) and $\mu/T=2.0$ (red dotted line). 
The value of the strange quark chemical potential is set to $\mu_s=0.2\mu$.
}
\label{Fig:QuarkMasses2}
\end{figure}

\subsection{Phase diagram}

In Fig.~\ref{Fig:Phasdig} we show the phase diagram of the present model.
To this end we find the positions of the minima of the temperature derivative (the steepest decrease) of the quark mass as the chiral order parameter $dM/dT$ in the $T-\mu$ plane and display them by the dashed line. These pseudocritical temperatures go over to the critical temperatures of the first order phase transition characterized by a jump of the quark mass at the corresponding position in the $T-\mu$ plane.

A characteristic feature of the phase diagram is that lowering the ratio $T/\mu \to 0$, the phase transition turns from crossover to first order. The chiral restoration is a result of the phase space occupation due to Pauli blocking which effectively reduces the interaction strength in the  gap equation.
\subsection{Mass spectrum for quarks and mesons at finite temperatures}
\begin{figure}[!th]
\centerline{\includegraphics[height=0.192\textheight, width=0.4\textwidth]{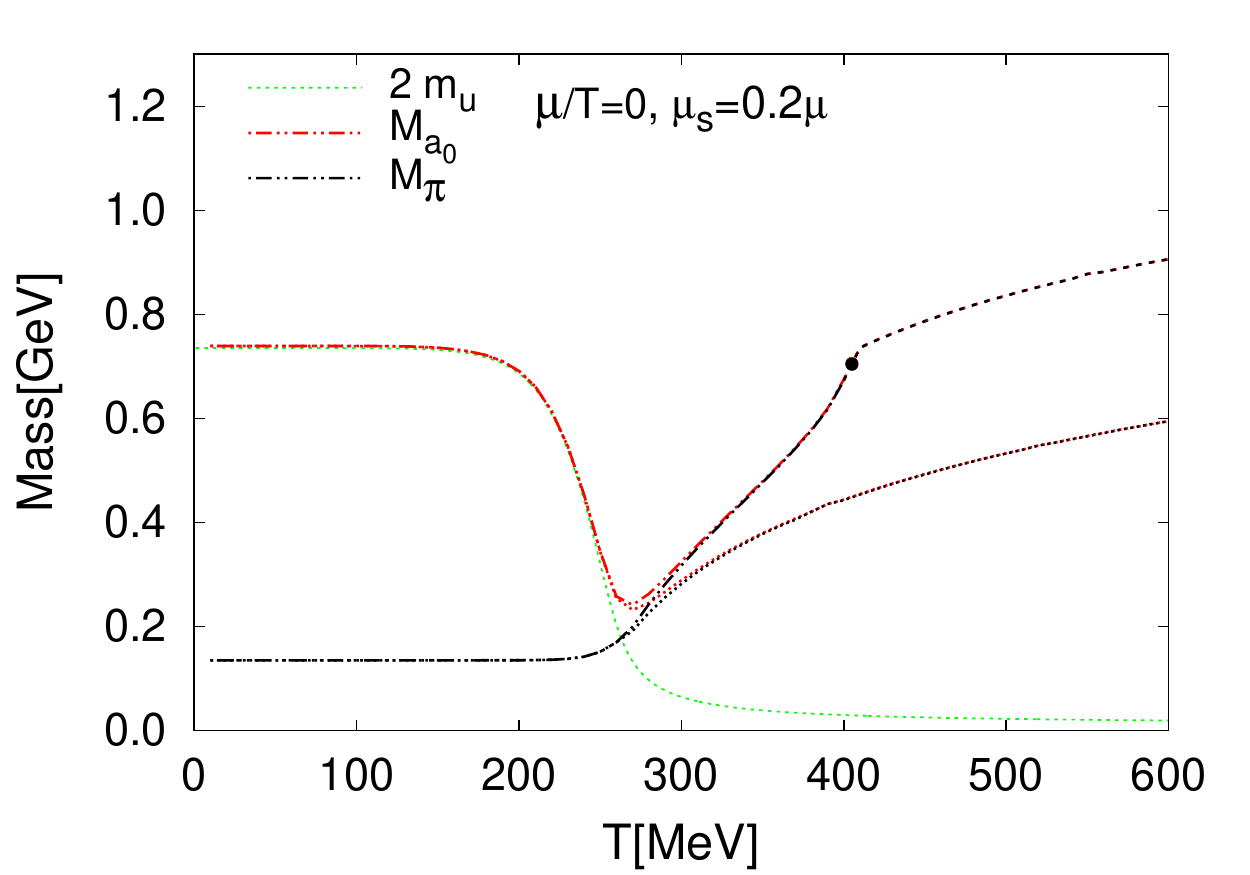}}\centerline{\includegraphics[height=0.192\textheight, width=0.4\textwidth]{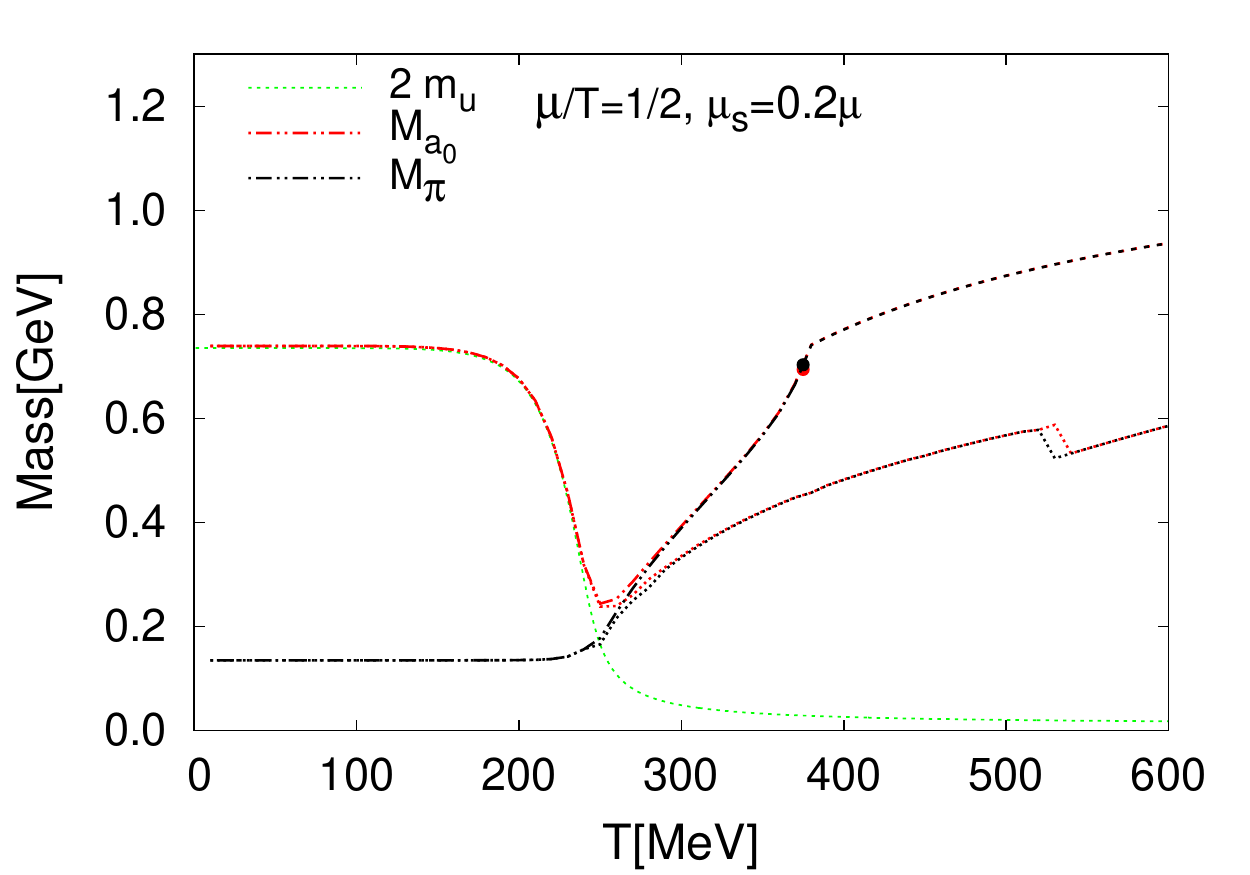}}
\centerline{\includegraphics[height=0.192\textheight, width=0.4\textwidth]{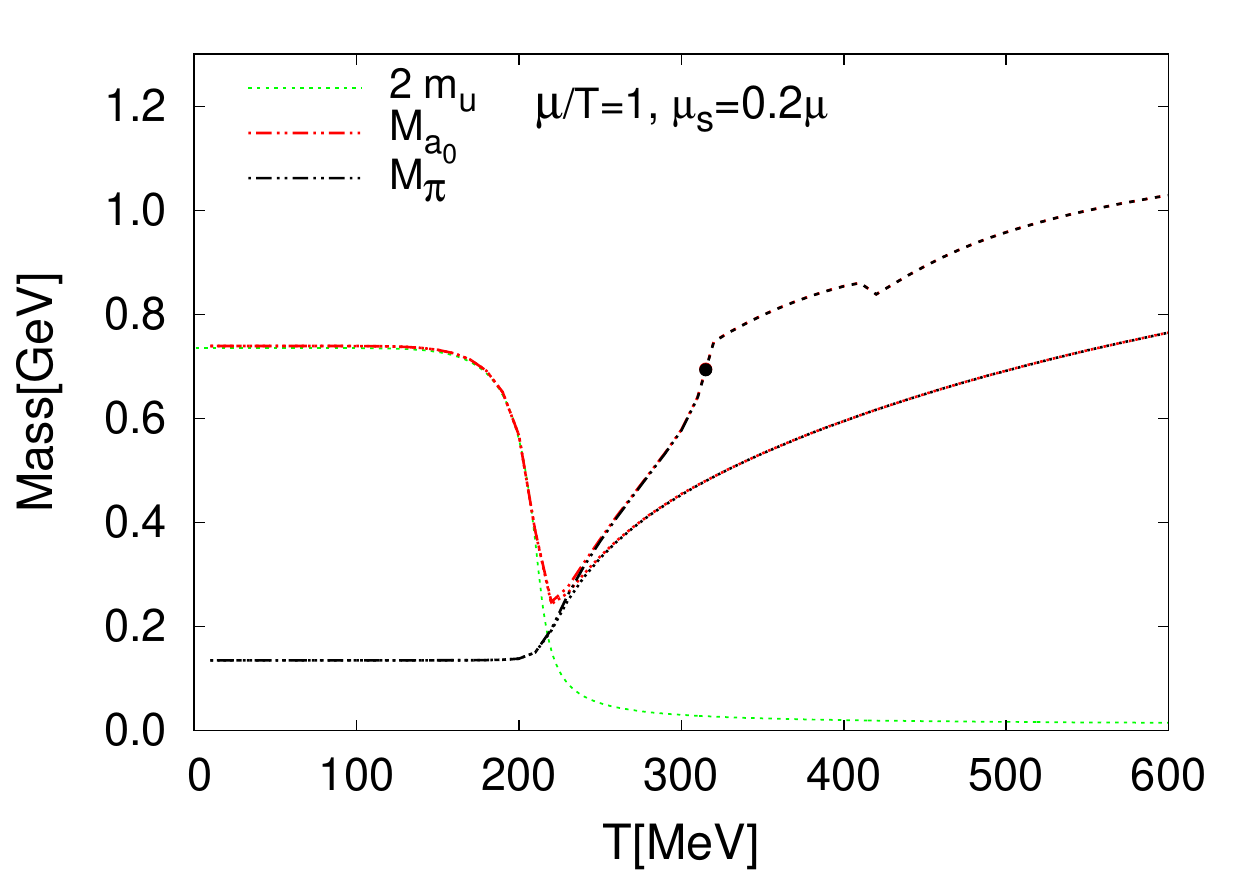}}
\centerline{\includegraphics[height=0.192\textheight, width=0.4\textwidth]{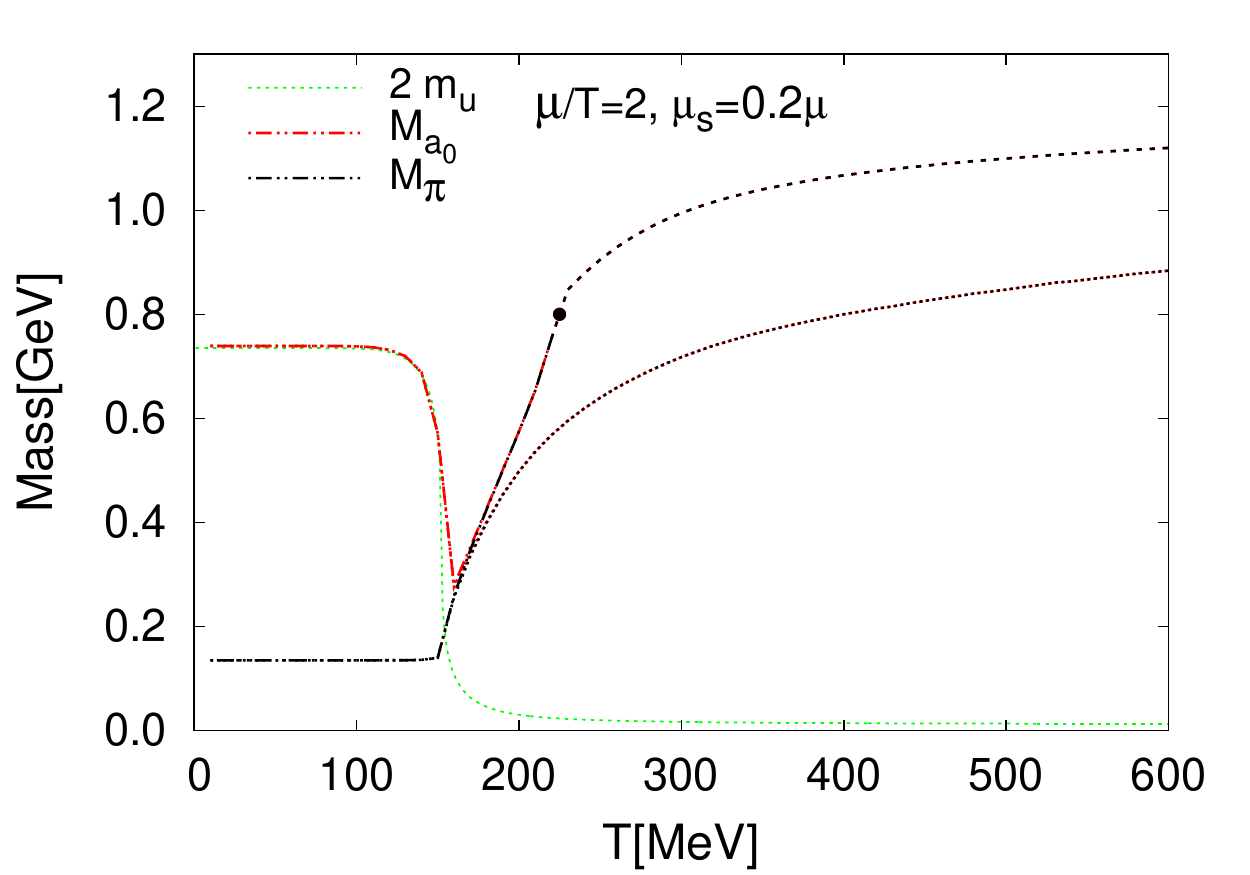}}
\caption{
Temperature dependence of the masses for pion (black dash double-dotted line) and $a_0$ meson (red dash double-dotted line) together with continuum thresholds $m_{\rm thr,\pi}=2 m_u$ (green dotted line)
for $\mu/T=0$, $0.5$, $1.0$ and $2.0$ (from top to bottom panel). 
Different line styles correspond to different mass definitions: as a maximum of the derivative of the phase shift (lower line to the right) or when the phase shift value hits $\pi/2$ (upper line). 
The point on the line denotes the temperature when the phase shift is lower than $\pi/2$ and in this case
the line after that point corresponds to the position of the maximum of the phase shift derivative.
\label{Fig:MesonMassPiMu0}}%
\end{figure}

\begin{figure}[!h]
\centerline{\includegraphics[height=0.222\textheight, width=0.4\textwidth]{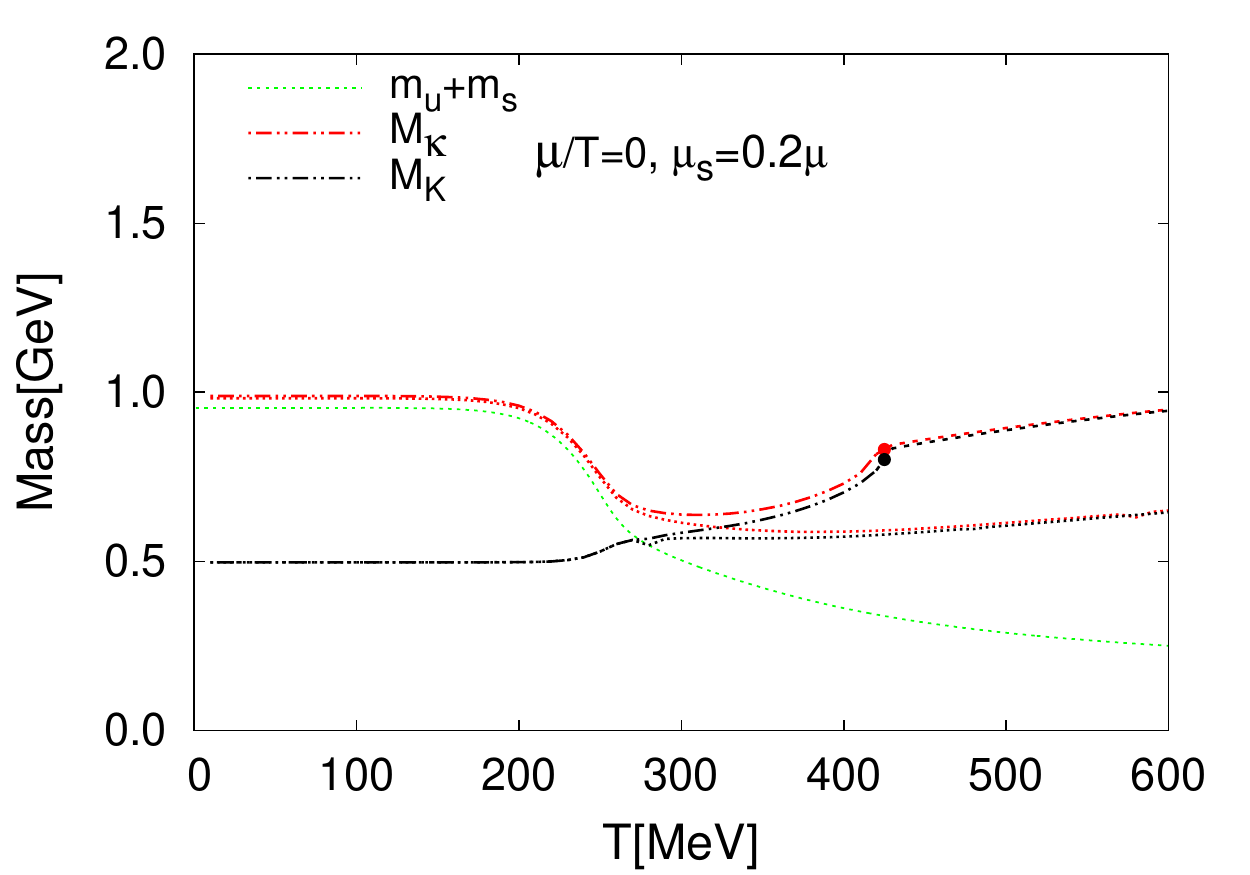}}
\centerline{\includegraphics[height=0.222\textheight, width=0.4\textwidth]{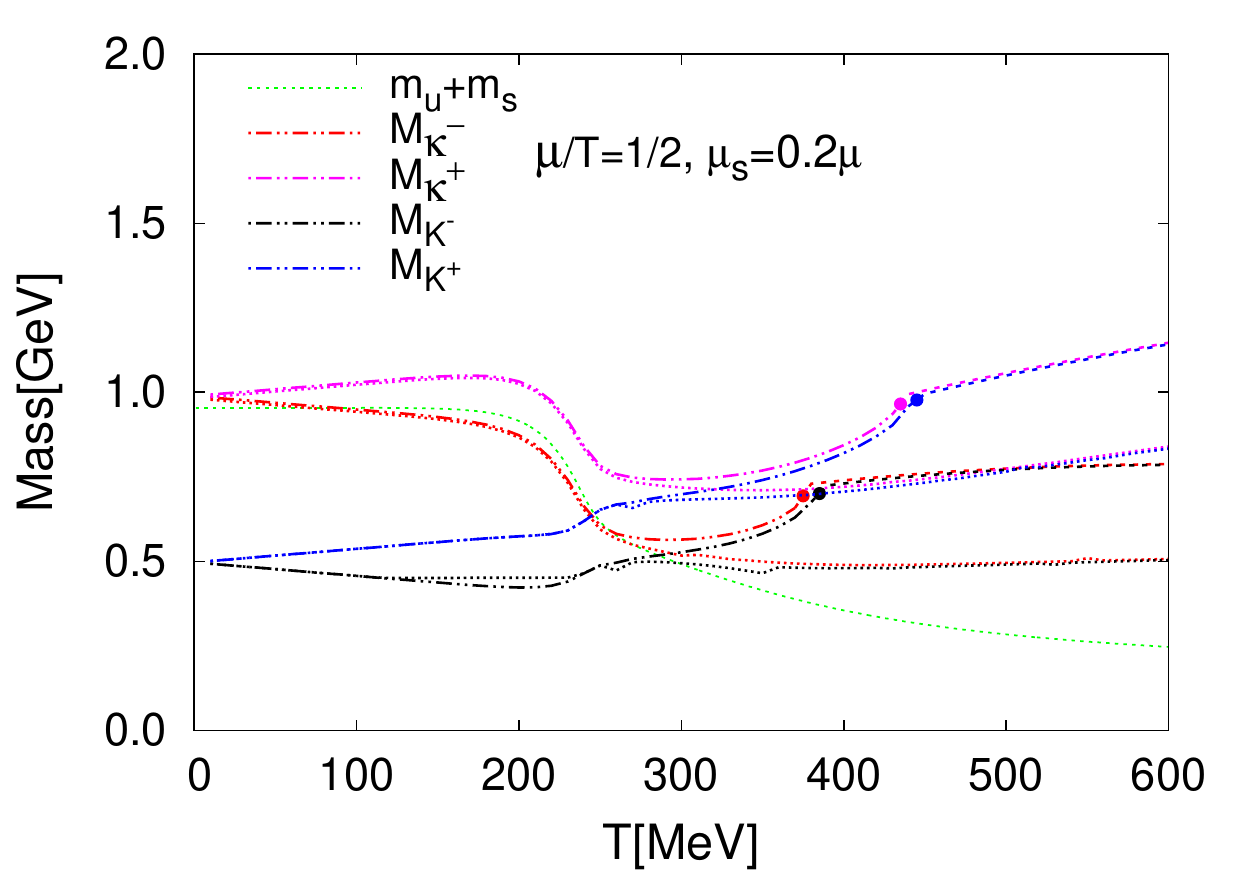}}
\centerline{\includegraphics[height=0.222\textheight, width=0.4\textwidth]{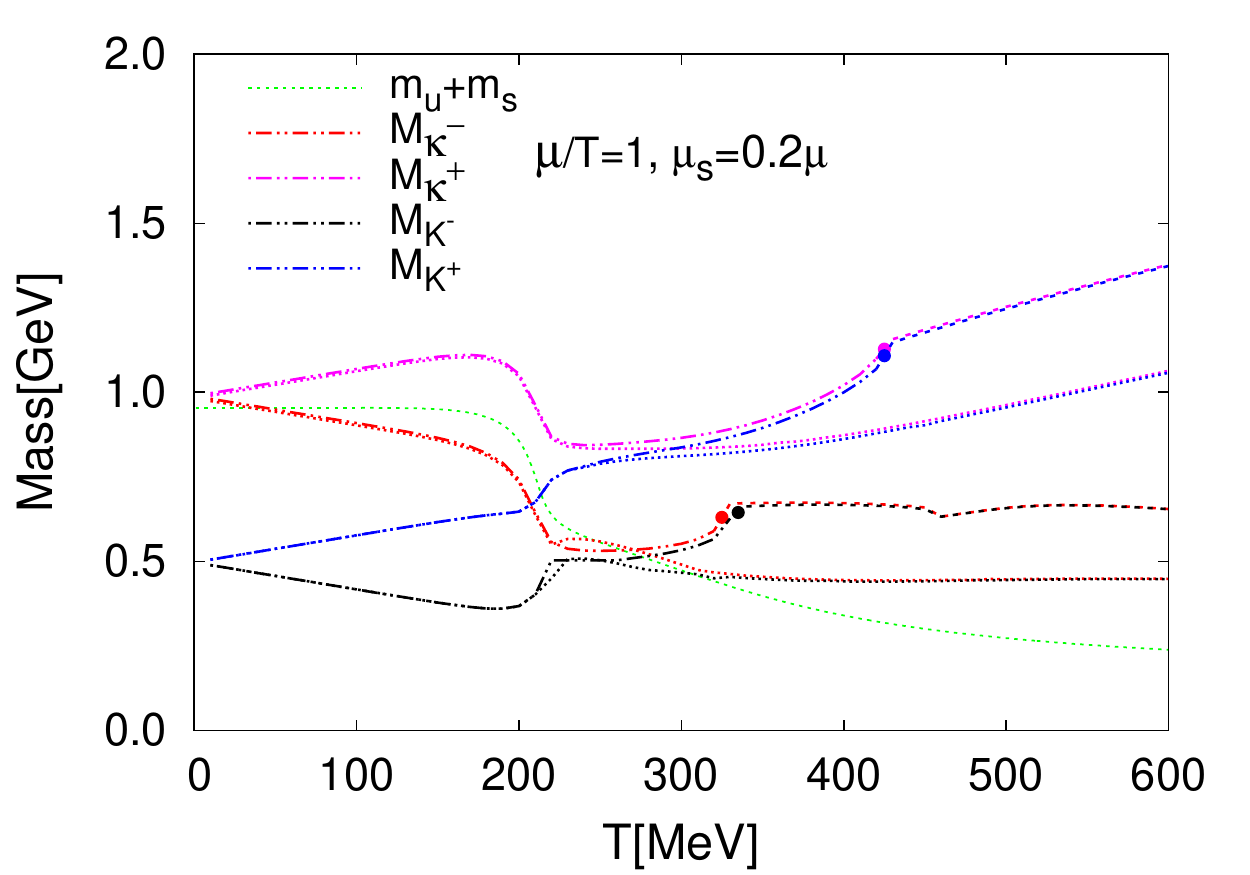}}
\centerline{\includegraphics[height=0.222\textheight, width=0.4\textwidth]{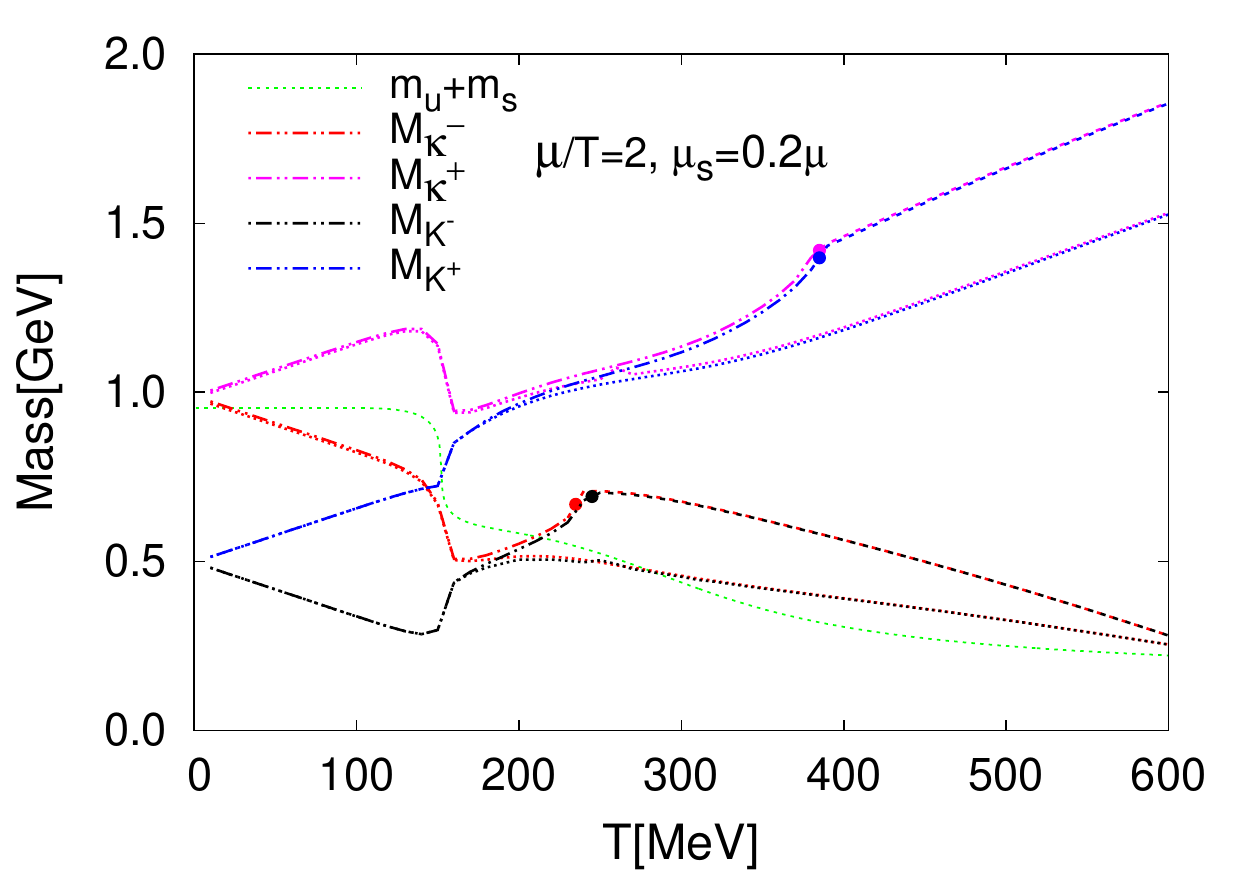}}
\caption{
Same as Fig.~\ref{Fig:MesonMassPiMu0}, but for kaons and $\kappa$ mesons.
The continuum threshold is given by $m_{\rm thr,\pi}=m_u + m_s$ (green dotted line).
\label{Fig:MesonMassKaMu01}}
\end{figure}

In Fig.~\ref{Fig:QuarkMasses2}  we show the masses of quarks together with the sum and the difference of the traced Polyakov loop $\Phi$ and its conjugate $\bar{\Phi}$ along different trajectories  
$\mu/T={\rm const}$ in the $T-\mu$ plane. 
We note that the chiral symmetry restoration in the light quark sector is correlated with a rise in the traced 
Polyakov loop which, due to the polynomial form of the Polyakov-loop potential, also exceeds the value 1.
For finite chemical potentials $\Phi \neq \bar{\Phi}$ holds in the domains where $\Phi\neq 0$.  
In Figs.~\ref{Fig:MesonMassPiMu0} and \ref{Fig:MesonMassKaMu01} we present masses of the pseudoscalar $\pi$ and $K$ mesons as well as of their scalar partners, the $a_0$ and $\kappa$ (kappa) mesons.  
These masses may be found in different ways.   

The first one is by solving the Bethe-Salpeter equation,  where the mass is obtained as the pole of the meson propagator. 
This method works well as long as the particle is a true bound state, that is below  the Mott temperature 
($T_{\rm Mott}^{M}$) for the meson $M$.   
Above the Mott temperature the meson becomes an unbound state and the definition of mass is complicated by the fact that the pole becomes complex and the solution is not unique.  
Therefore, to find  approximate solutions in this case one is generally using the  Breit-Wigner form of the propagator with the width defined by $\Gamma={\rm Im}(q_0)$.  
However, this approximation  is valid only close to $T_{\rm Mott}^{M}$ as it was discussed in our previous work \cite{Blaschke:2013zaa}. 

The second method of finding the meson masses involves  the phase shift of the quark-antiquark correlation in the considered mesonic interaction channel. 
To define the mass we determine the energy $\omega$ where the phase shift assumes the value $\pi/2$.
In the rest frame of the meson this energy corresponds to the mass. 
Below the Mott temperature, the phase shift jumps from zero to $\pi$ at this position so that its derivative 
is a delta function, characteristic for a true bound state.  
We do not consider the positive mode at low energy which arises in the strange sector for 
${K}$, ${\kappa}$. 
When the phase shift stays below $\pi/2$, the "mass" can be defined by the position of the maximum  
of the phase shift. 
For the behaviour of the phase shifts see Figs.~\ref{Fig:ShiftsPiSiV} - \ref{Fig:ShiftsKapKV1}.    

The third method to determine the meson mass also involves the phase shift.  
One finds the value of the mass from the position of the maximum of the derivative of the phase shift. 
For a true bound state, the derivative of the phase shift  corresponds to a delta function 
while for a resonance in the continuum it is a smooth function \cite{Blaschke:2013zaa,Wergieluk:2012gd}. For a meson at rest in the medium, the position of the peak of the energy derivative of the phase shift corresponds to the value of the meson mass.

The finite chemical potential in the case nonzero $\mu/T$  removes the degeneracy of the meson masses
in the strange channels. 
A mass difference arises between $K^{+}$ and $K^{-}$ as well as between their chiral partners 
$\kappa^{+}$ and $\kappa^{+}$ see Fig.~\ref{Fig:MesonMassKaMu01}. 
The chemical potential shifts the pole in the propagator, which results in a reduction of the pseudocritical temperature $T_{c}$ and therefore also in a reduction of the meson Mott temperatures $T_{\rm Mott}^{M}$.  

\subsection{Phase shifts for mesons and the Beth-Uhlenbeck equation of state}

\begin{figure}[!th]
\includegraphics[height=0.212\textheight, width=0.38\textwidth]{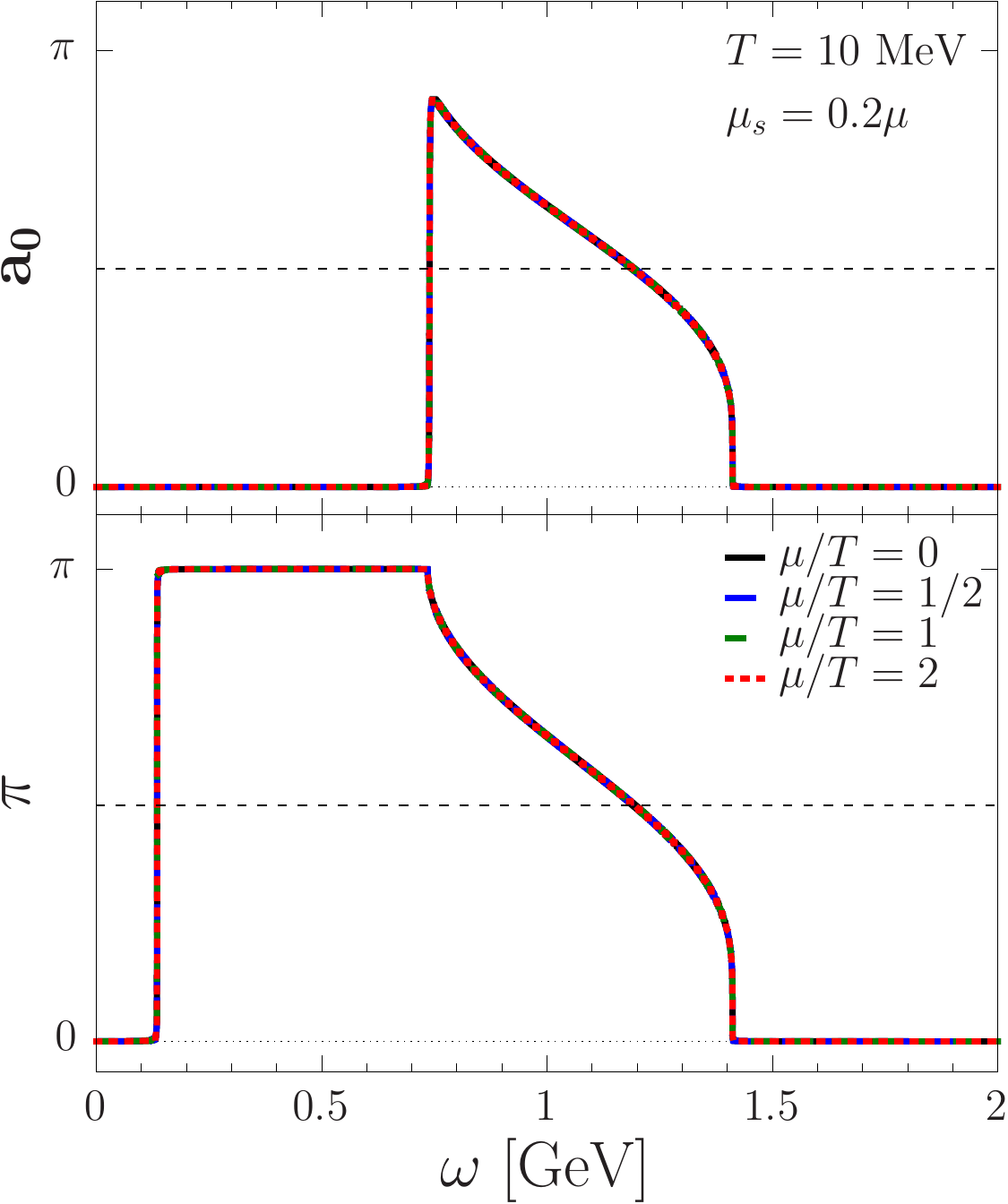}
\includegraphics[height=0.212\textheight, width=0.38\textwidth]{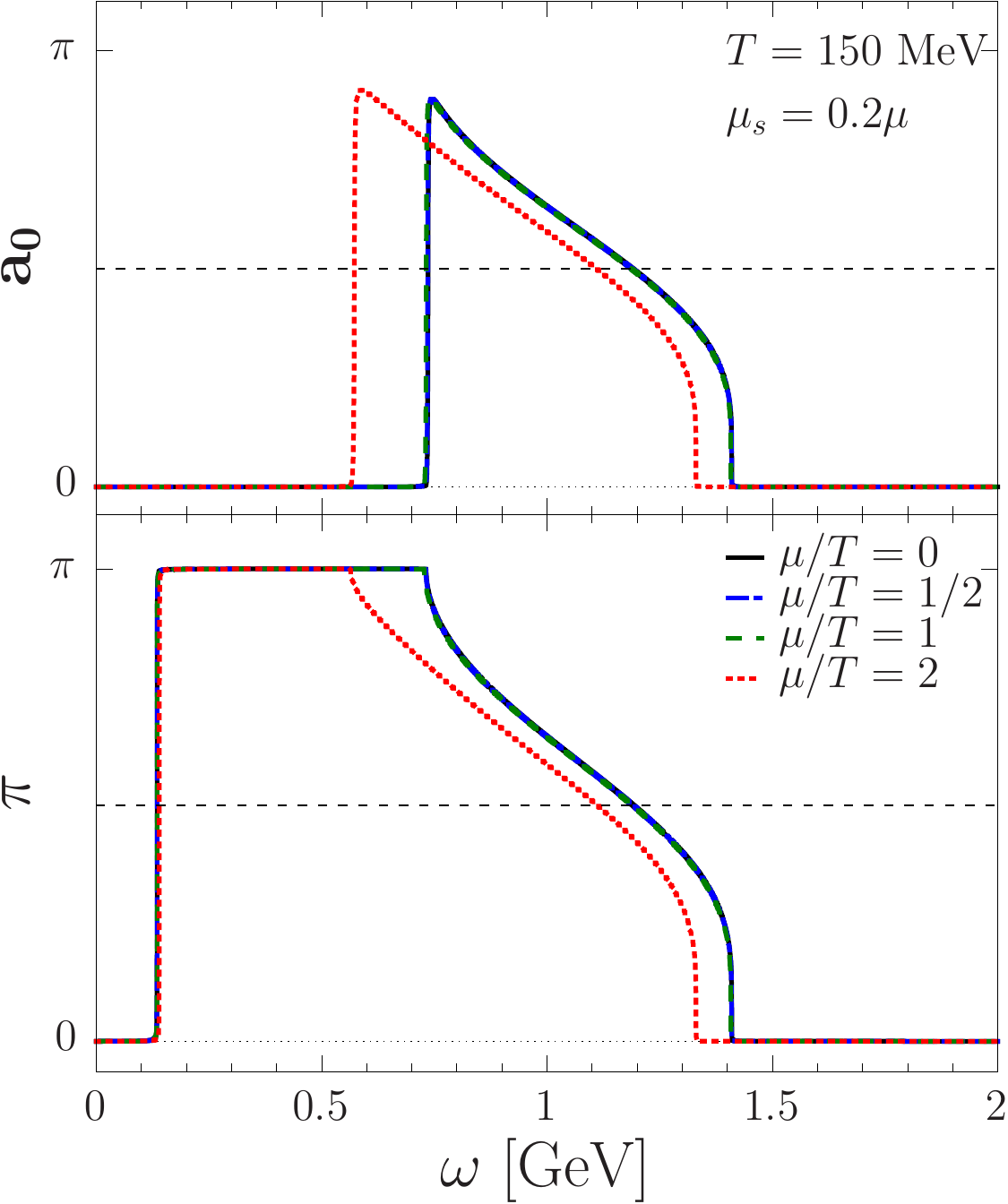}
\includegraphics[height=0.212\textheight, width=0.38\textwidth]{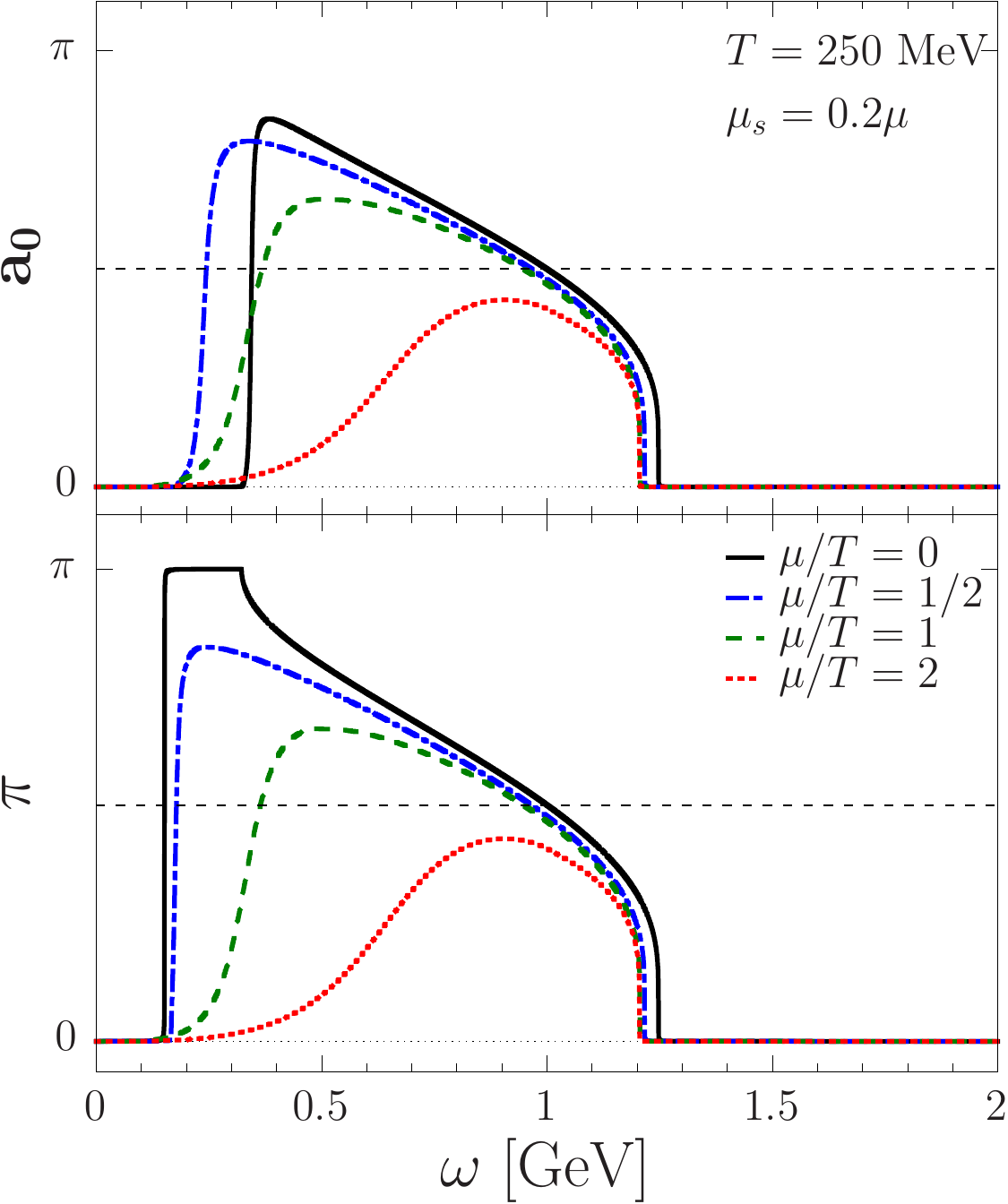}
\includegraphics[height=0.212\textheight, width=0.38\textwidth]{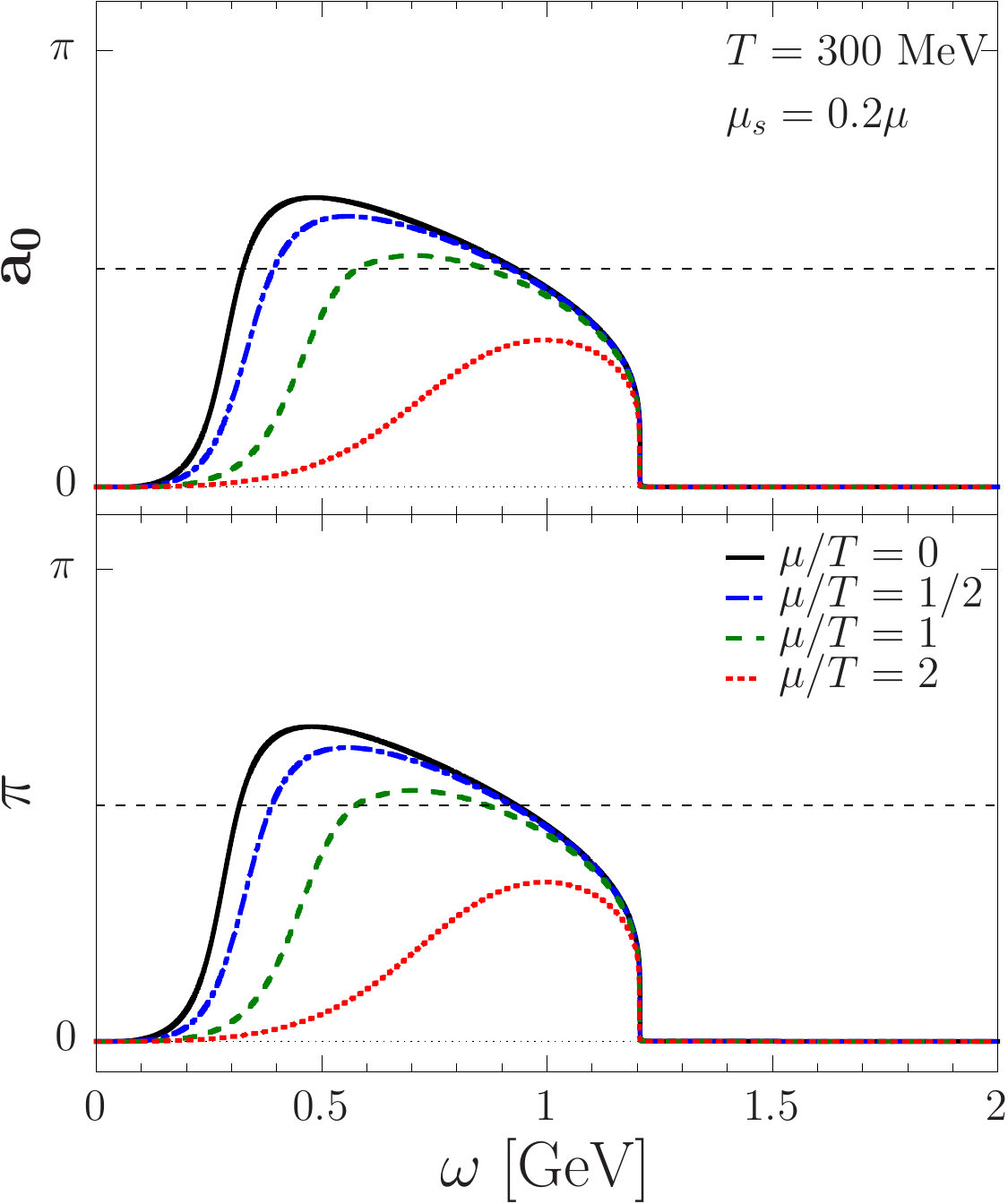}
\caption{Dependence of the phase shift in the pion and $a_0$ meson channel on the center of mass energy for different temperatures $T= 10$, $150$, $250$, $300$ MeV (from top to bottom) along different trajectories in the phase diagram: $\mu/T=0$ (black solid line), $\mu/T=0.5$ (blue dash-dotted line), $\mu/T=1.0$ (green dashed line) and $\mu/T=2.0$ (red dotted line).
}
\label{Fig:ShiftsPiSiV}%
\end{figure}

\begin{figure}[!th]
\includegraphics[height=0.23\textheight, width=0.38\textwidth]{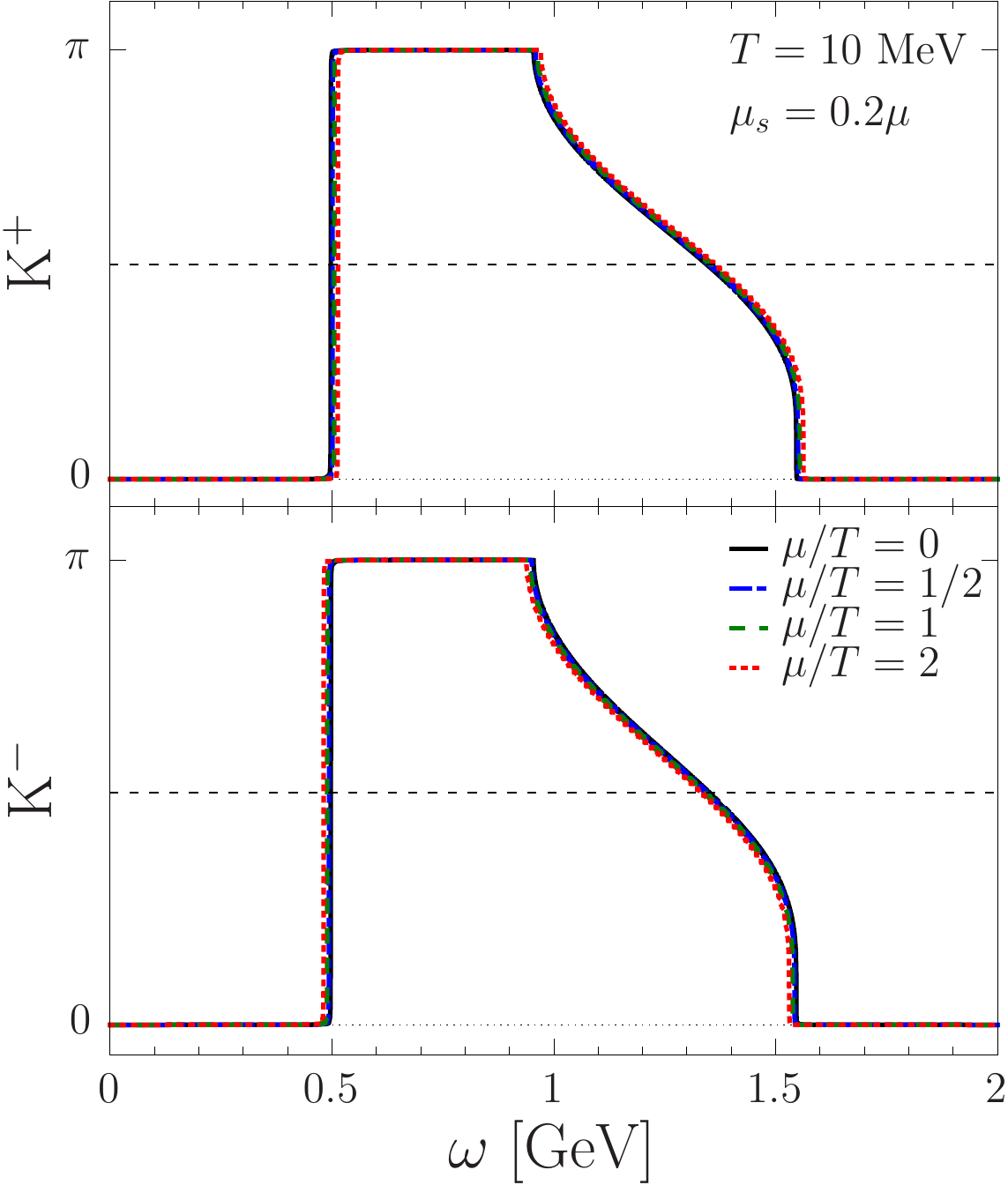}
\includegraphics[height=0.23\textheight, width=0.38\textwidth]{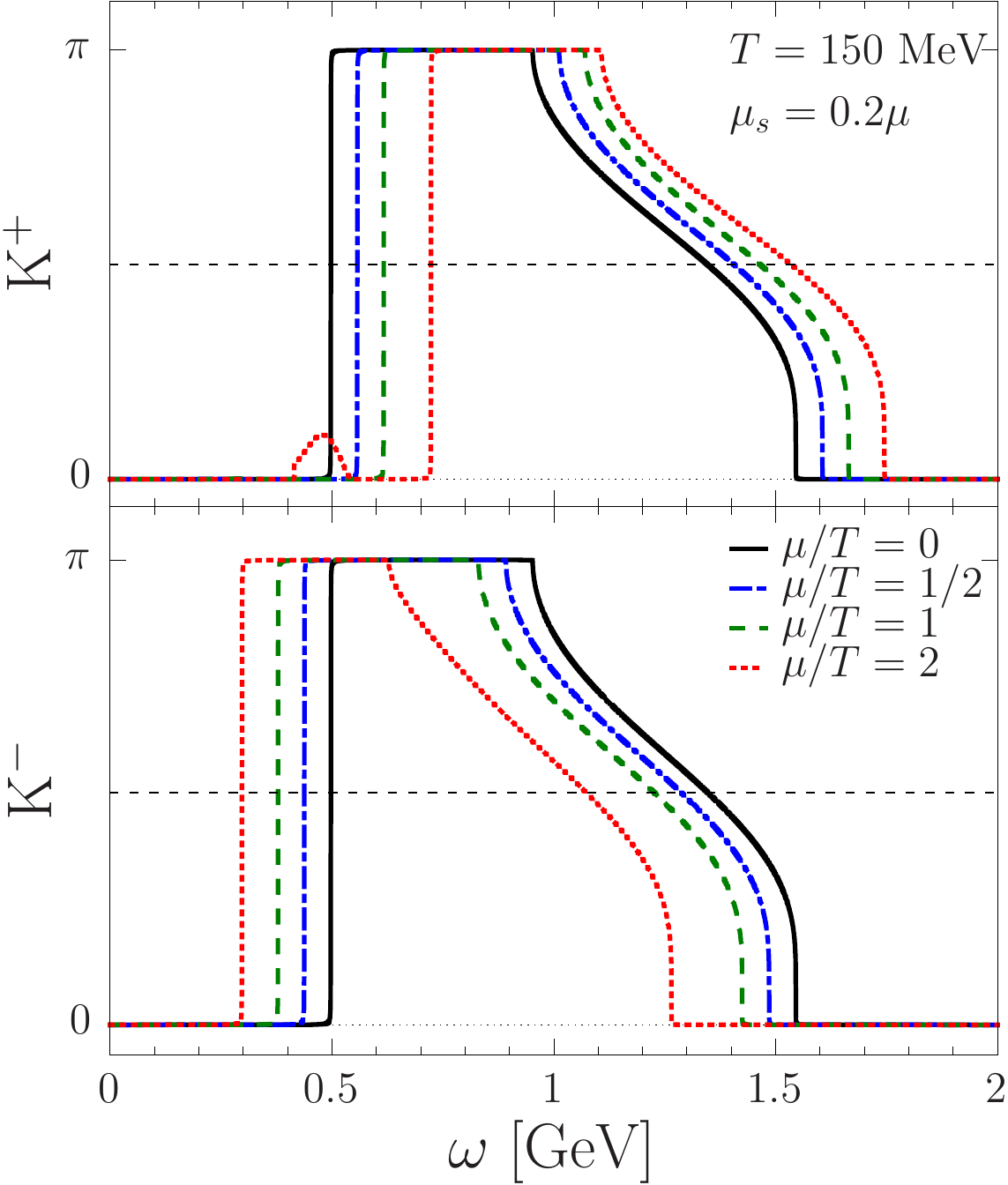}
\includegraphics[height=0.23\textheight, width=0.38\textwidth]{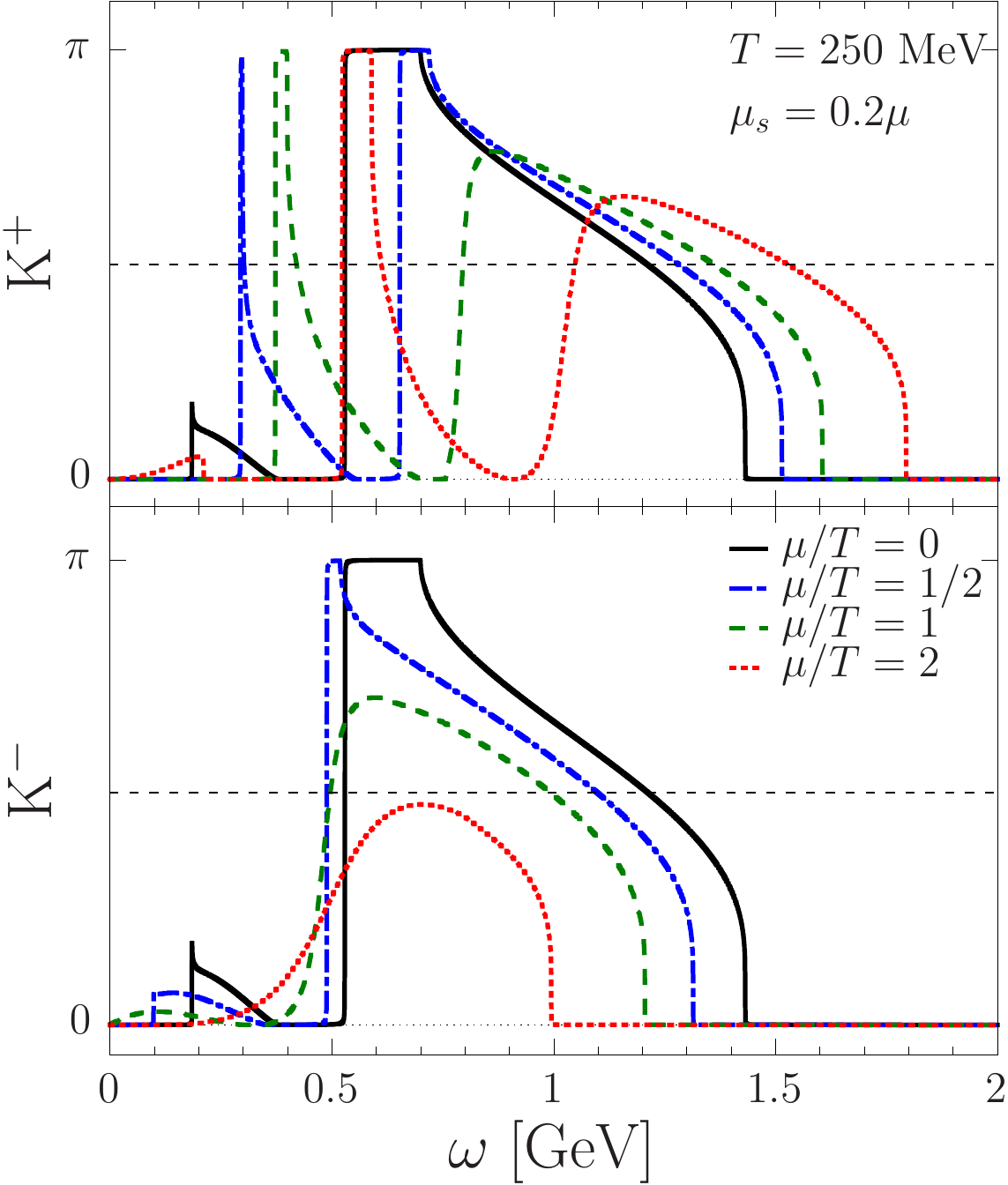}
\includegraphics[height=0.23\textheight, width=0.38\textwidth]{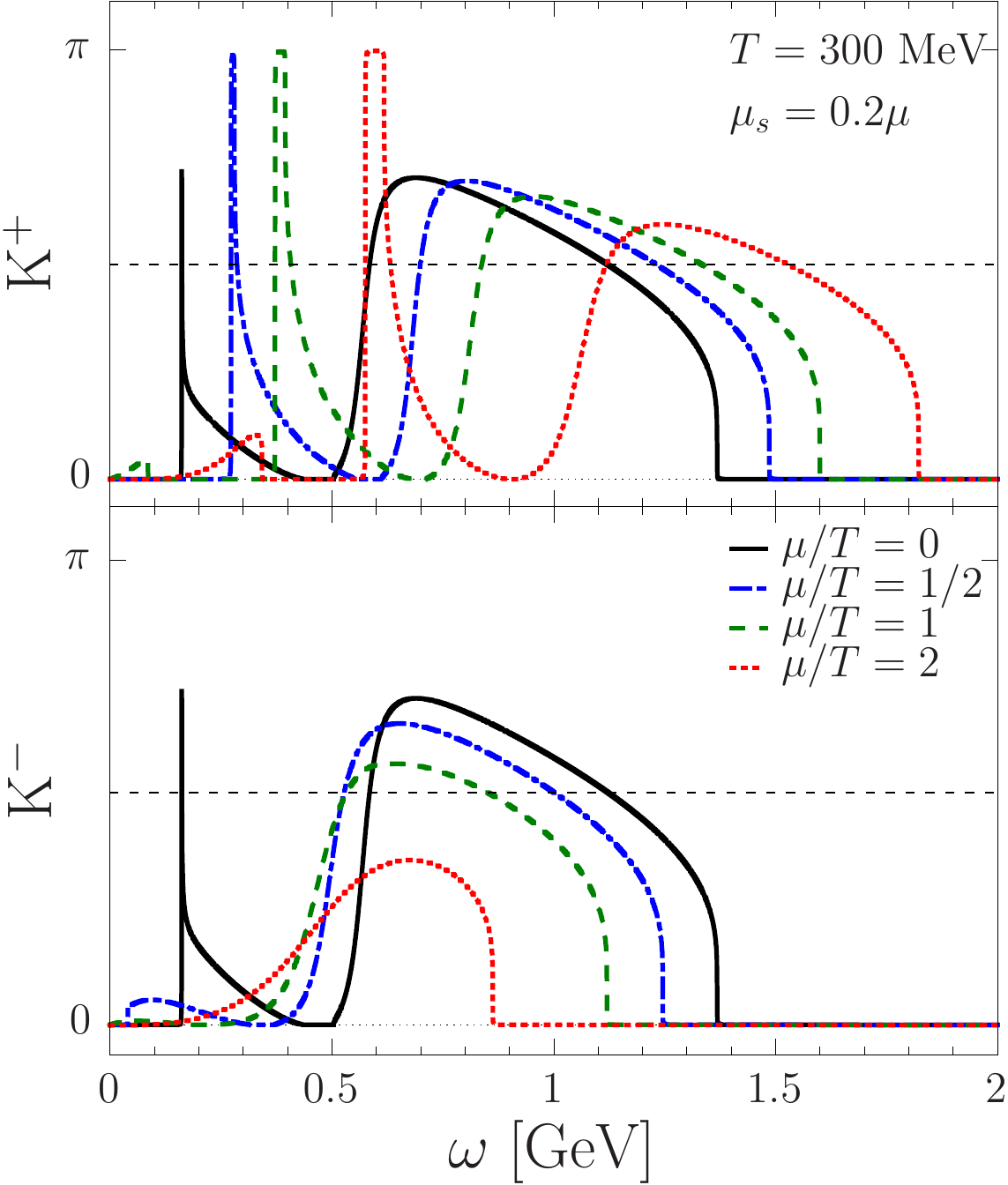}\\
\caption{
Same as Fig.~\ref{Fig:ShiftsPiSiV} for the $K^+$ and $K^-$ states.
}
\label{Fig:ShiftsKaKaV}
\end{figure}

\begin{figure}[!th]
\includegraphics[height=0.23\textheight, width=0.38\textwidth]{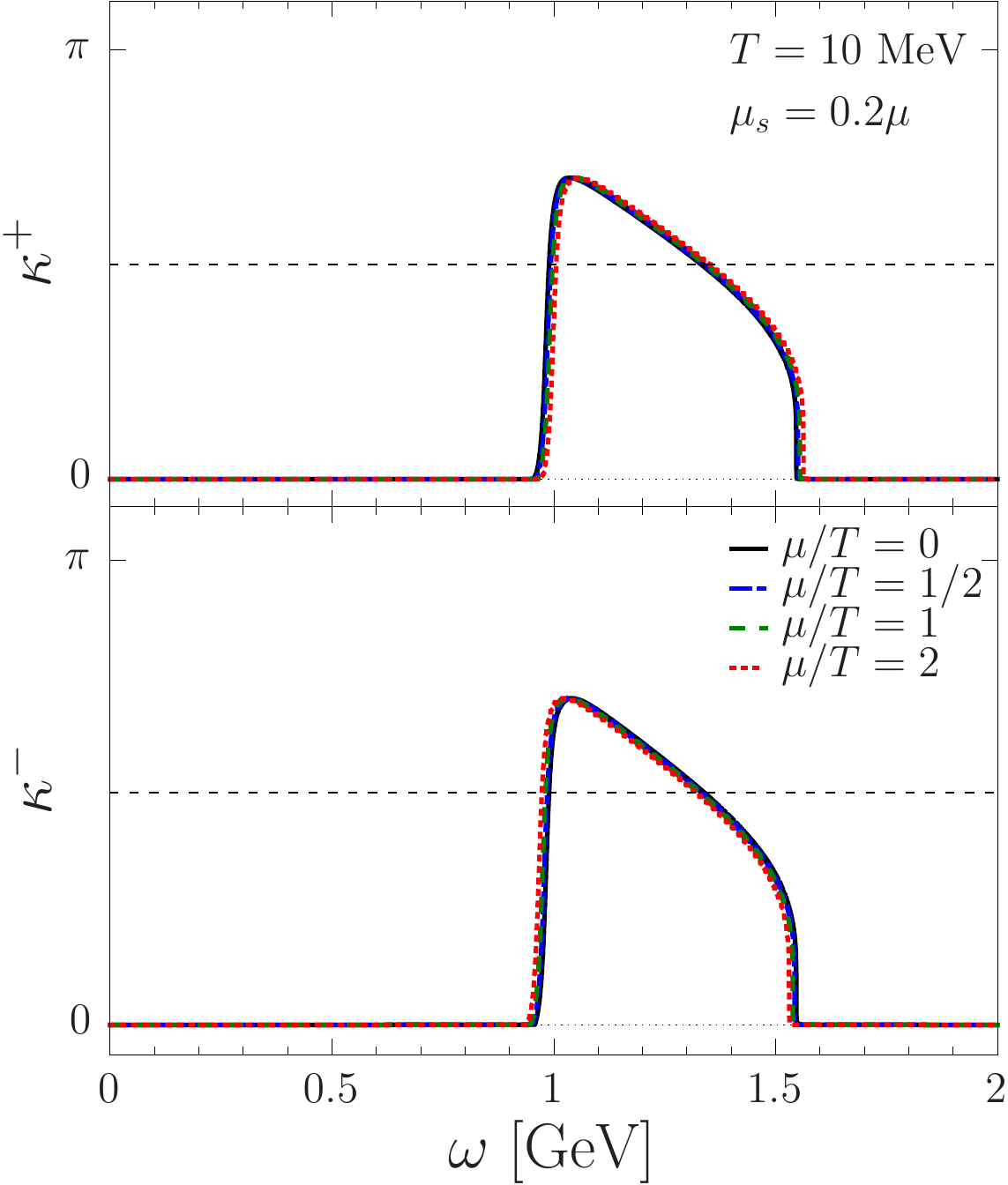}
\includegraphics[height=0.23\textheight, width=0.38\textwidth]{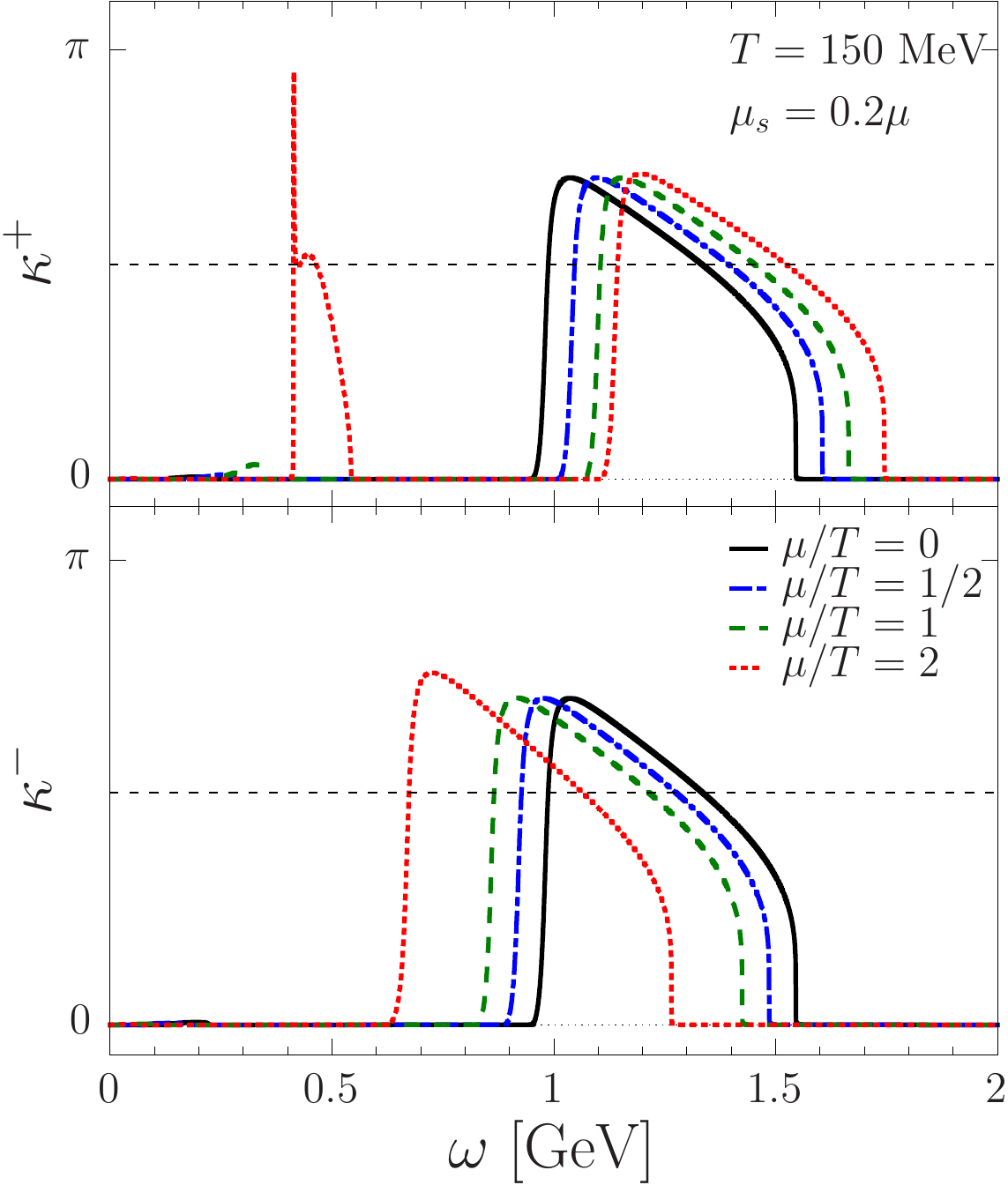}
\includegraphics[height=0.23\textheight, width=0.38\textwidth]{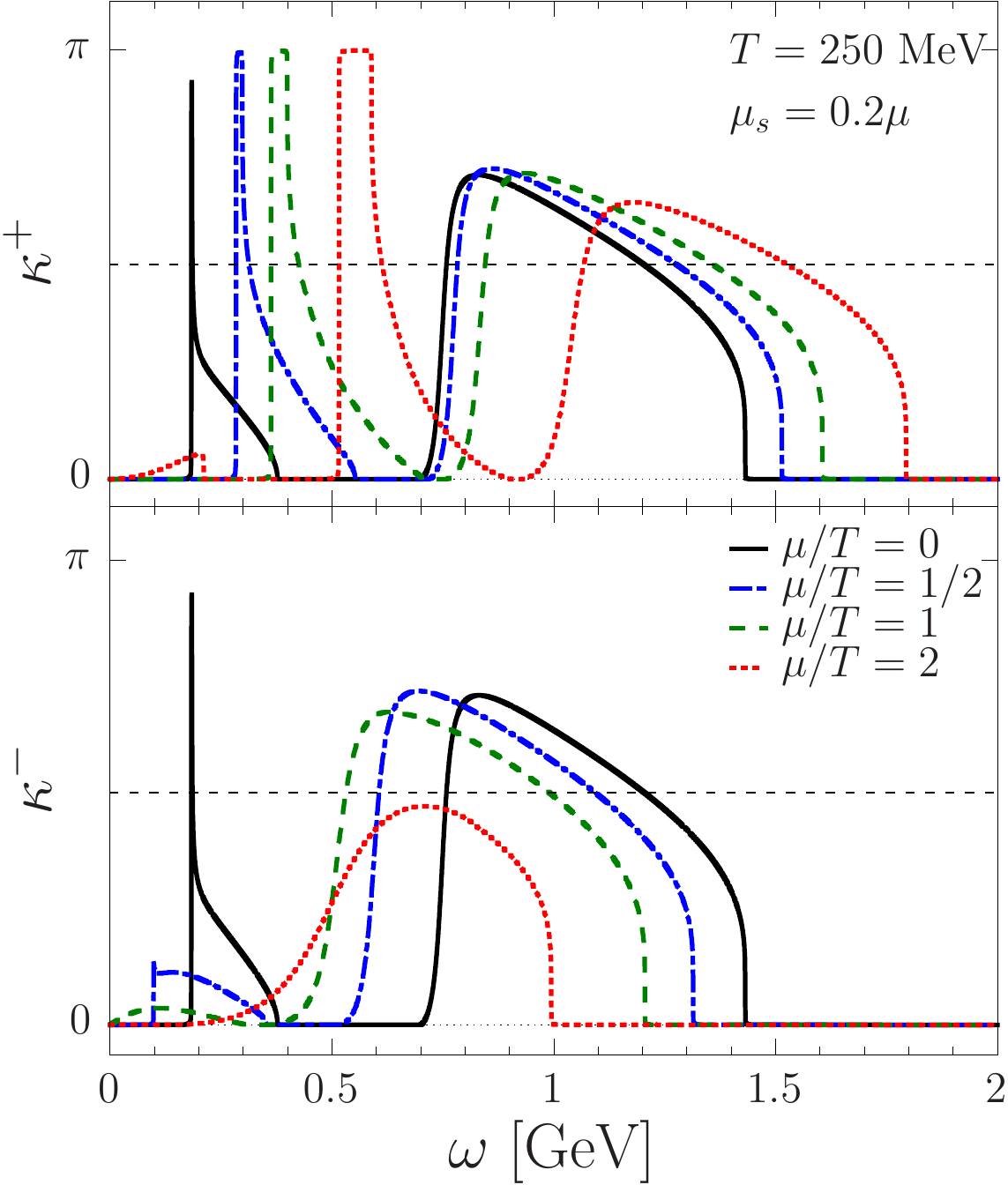}
\includegraphics[height=0.23\textheight, width=0.38\textwidth]{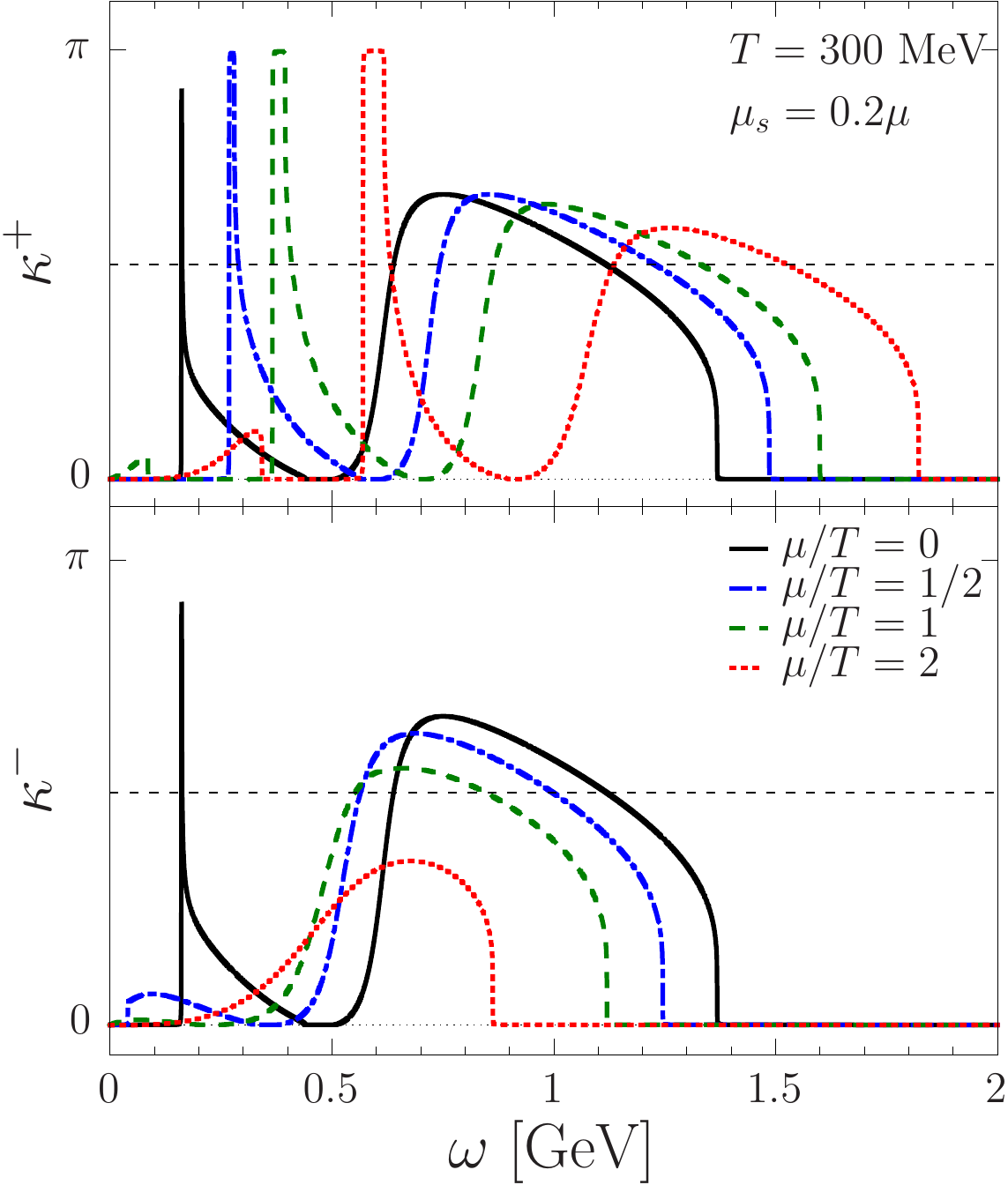}
\caption{
Same as Fig.~\ref{Fig:ShiftsKaKaV} for the $\kappa^+$ and $\kappa^-$ states.
}
\label{Fig:ShiftsKapKV1}
\end{figure}

In this subsection we discuss the results for the phase shifts for $\pi$ and $K$ mesons in comparison with those of their chiral partners, the $a_0$ and the $\kappa$ mesons, resp. (see also Ref.~\cite{Dubinin:2015glr}) and their consequences for the thermodynamics of quark-meson matter at finite temperature, 
with the coupling to the Polyakov-loop.  
The solutions for the meson phase shifts at finite temperature  are shown in Figs.~\ref{Fig:ShiftsPiSiV} - \ref{Fig:ShiftsKapKV1}  as  functions of the energy $\omega$ for different temperatures and values of the ratio $\mu/T$.  
As in Ref.~\cite{Blaschke:2013zaa} we have made the simplifying assumption that, even in the medium, 
the phase shifts are Lorentz invariant and depending on $\omega$ and $\bf q$ only via the Mandelstam variable $s=\omega^{2}-{\bf q}^2$ in the form $\delta_{\rm M}(\omega,{\bf q})=\delta_{\rm M}(\sqrt{s},{\bf q}=0)\equiv\delta_{\rm M}(\sqrt{s};T,\mu_{\rm M})$ for given temperature and chemical potential of the medium. 
Then the BU formula for the mesonic pressure can be given the following form 
\begin{eqnarray}
\label{GBU_M}
P_{\rm M}
	&=&
	d_{\rm M} \int\frac{{\rm d}^3q}{(2\pi)^3}~
	\int_0^\infty\frac{{\rm d}s}{4 \pi}~
	\frac{1}{\sqrt{s+q^2}}\bigg\{ 
	g(\sqrt{s+q^2}-\mu_{\rm M})
\nonumber\\
&&+g(\sqrt{s+q^2}+\mu_{\rm M})
	\bigg\}
	\delta_{\rm M}(\sqrt{s};T,\mu)
	~.
	\label{BUU}
\end{eqnarray}
The bound state mass is located at the jump of the phase shift from $0$ to $\pi$ and this jump corresponds to a delta-function in the BU formulas (\ref{BUU}) for the correlations.   
In the case when the continuum of the scattering states can be neglected, that is when it is separated by a sufficiently large energy gap from the bound state, we obtain as a limiting case the thermodynamics of a statistical ensemble of on-shell correlations (resonance gas).  

In Fig.~\ref{Fig:ShiftsPiSiV} we show the phase shifts for the nonstrange mesons $\pi$ and $a_0$.
As expected for pseudoscalar and scalar isovector states, their phase shifts reflect the chiral symmetry 
breaking at low temperatures (and low chemical potentials), with a deeply bound pion and an $a_0$ state just above the continuum threshold, separated from the pion by a large gap.  
At high temperature $T=300$ MeV  these phase shifts become indistinguishable, reflecting the chiral symmetry restoration.

In Fig.~\ref{Fig:ShiftsKaKaV} we show the phase shifts for the charged kaons. For $\mu=0$ their modification with increasing temperature parallels that of the pions, with the gap between the bound state and the continuum diminishing with temperature and becoming zero for $T=300$ MeV, above the kaon Mott temperature, where the kaon becomes a resonance in the continuum. 
At finite chemical potentials the opposite charge kaons develop a mass splitting and a new low-energy mode appears in the spectrum due to the finite mass difference, see also \cite{Yamazaki:2013yua}.  

In Fig.~\ref{Fig:ShiftsKapKV1} the phase shifts for the chiral partner states of the kaons, the charged 
$\kappa$ mesons, are shown. 
Comparing these results with those for the kaons parallels the comparison of the scalar $a_0$ meson 
with the pion.
The mass splitting of the opposite charge states with increasing chemical potential mirrors the behavior of the kaons. At low temperatures the chiral symmetry breaking is manifest in the mass splitting and energy gap between the kaons and the $\kappa$ mesons as chiral partner states. 
At high temperatures, when chiral symmetry is restored, these differences vanish and kaon and 
$\kappa$ phase shifts become indistinguishable.

Now we want to study the thermodynamics of mesons in a hot and dense medium encoded in the thermodynamic potential or equivalently the pressure (\ref{BUU}). 
Eq.~(\ref{BUU}) has the BU 
form, where the phase shift obeys the Levinson theorem in a medium \cite{Wergieluk:2012gd,Dubinin:2013yga}      
\begin{eqnarray}
\label{Levinson}
\int^{\infty}_{0}d\omega\frac{d\delta_{\rm M}(\omega)}{d\omega}~=~0.
\end{eqnarray}

Here we may introduce the energy level of the continuum threshold $\omega_{\rm thr}$ and split 
(\ref{Levinson}) in the sum of two integrals with the intervals of integration  $[0,\omega_{\rm thr}]$ and $[\omega_{\rm thr},\infty]$, respectively.  
After integrating out  we obtain the Levinson theorem in the form \cite{Blaschke:2014zsa}
\begin{eqnarray}
\label{Levinson2}
\pi ~n_{\rm B,M} = \delta_{\rm M}(\omega_{\rm thr}) - \delta_{\rm M}(\infty)~,
\end{eqnarray}
which applies also in the case of a hot and dense medium. 
The continuum threshold is $\omega_{\rm thr} =\sqrt{q^2+m_{\rm thr}^2}$,  where  $m_{\rm thr}=2m_u$ for the light-light quark mesons ($\pi,a_0$) and $m_{\rm thr}=m_u+m_{s}$ for the mixed light-strange two-particle states (${K},{\kappa}$). 
For energies below the threshold there can be only a discrete number $n_{\rm B,M}$ of bound states in the channel ${\rm M}$, each contributing an amount of $\pi$ to the change in the phase shift at the bound state energies $\omega_{\rm M, i}=\sqrt{q^2+M_{\rm i}^2}$ with ${\rm i}=1,\dots, n_{\rm B,M}$.
In particular, when due to the chiral symmetry restoration the dropping quark masses entail a lowering of the continuum thresholds $\omega_{\rm thr}$ this triggers the dissolution of the bound states into the continuum (the Mott effect) so that $n_{\rm B,M}=0$ results. 

\begin{figure}[!th]
\includegraphics[height=0.2\textheight, width=0.4\textwidth]{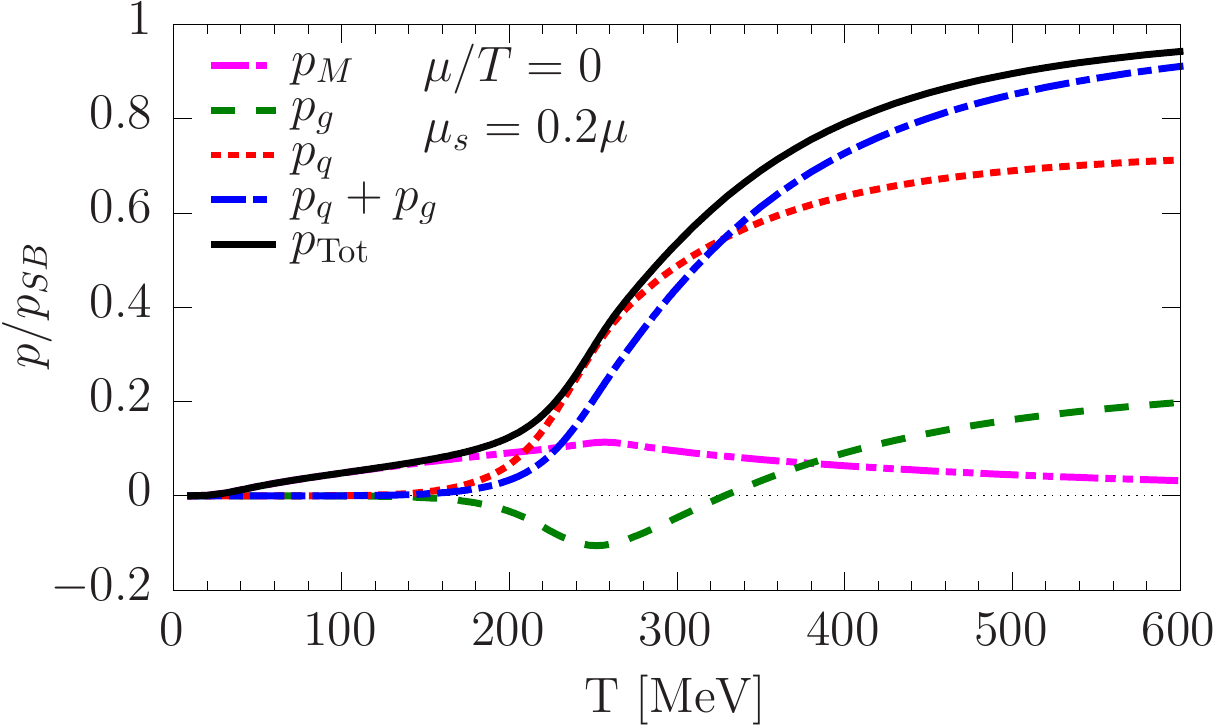}
\includegraphics[height=0.2\textheight, width=0.4\textwidth]{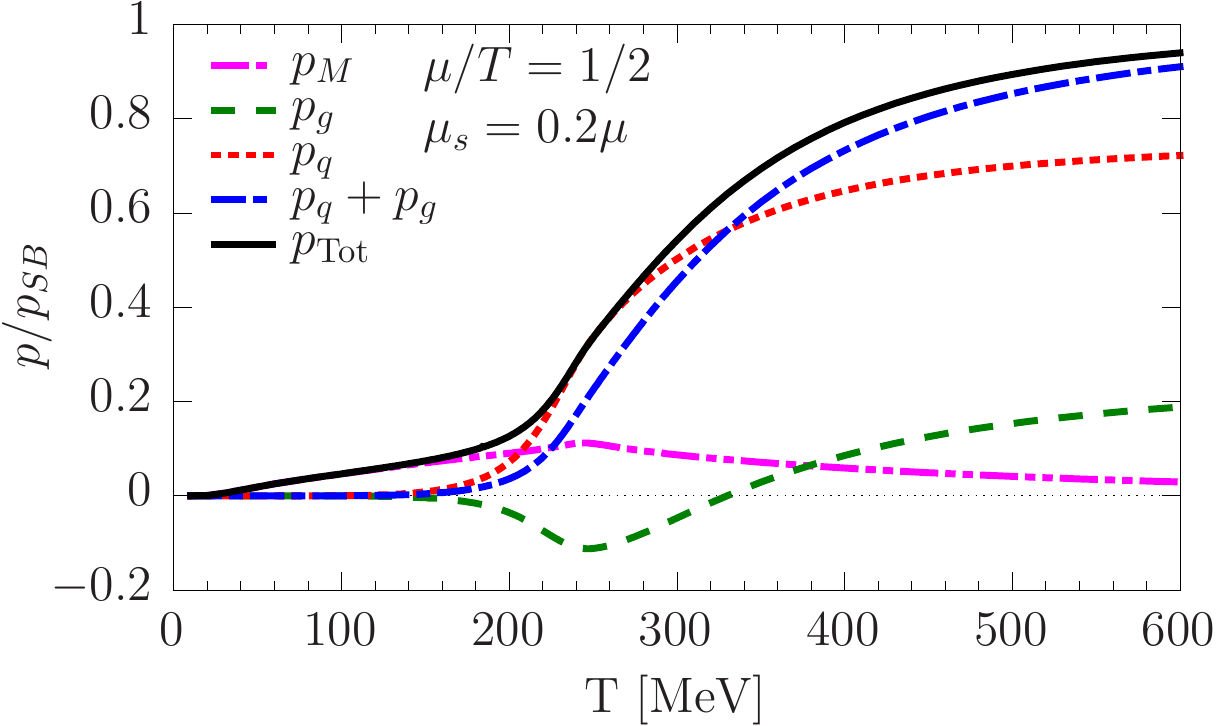}
\includegraphics[height=0.2\textheight, width=0.4\textwidth]{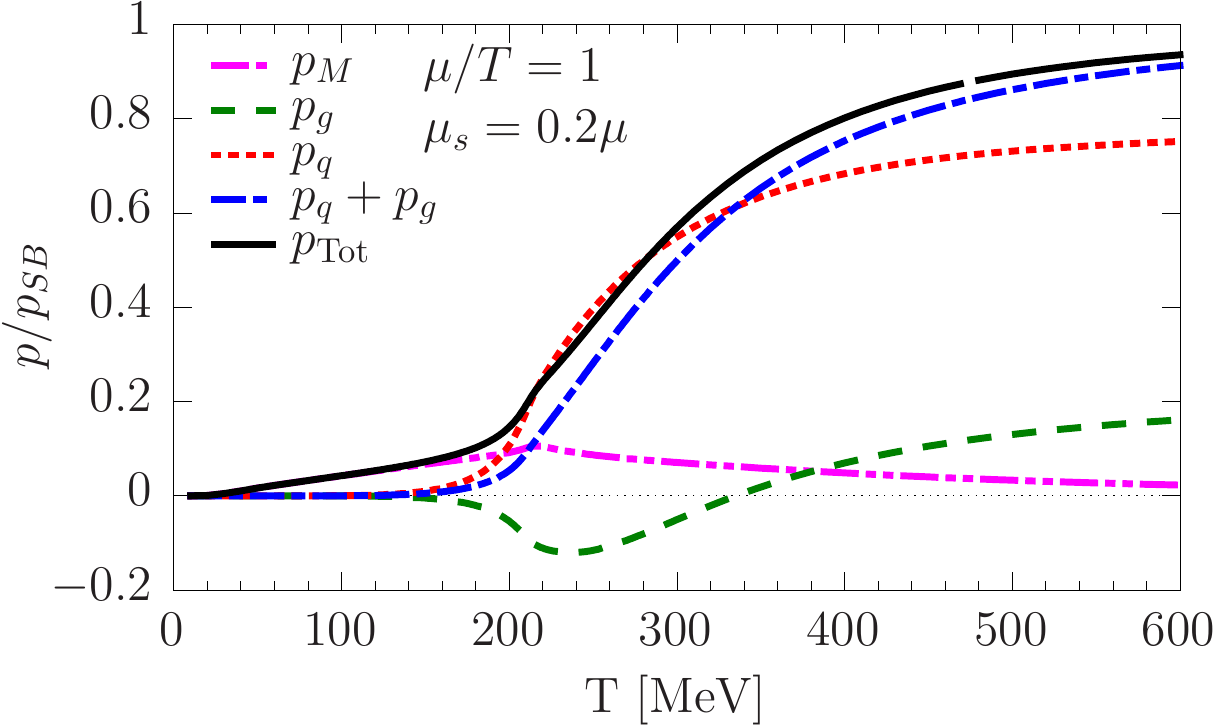}
\includegraphics[height=0.2\textheight, width=0.4\textwidth]{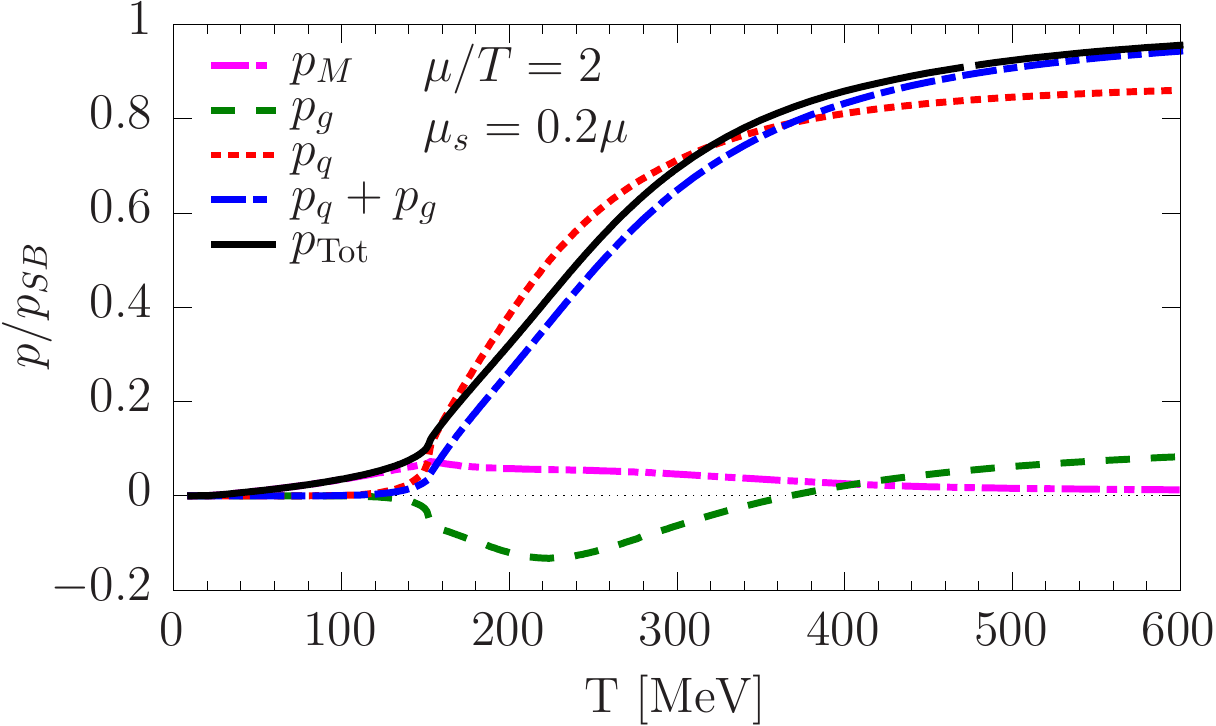}
\caption{Pressure of the $2+1$ flavor PNJL model with scalar and pseudoscalar meson correlations as a function of temperature (black solid line) normalized to the Stefan-Boltzmann pressure (\ref{P_SB}) for the cases $\mu/T=0, 0.5,1.0, 2.0$ (from top to bottom).  Also shown are the partial pressures of the components: light and strange quarks (red dotted); 'gluon' contribution from Polyakov loop potential $\mathcal{U}$ (green dashed);  light and strange quarks plus gluons (blue dashed-dotted); mesonic contributions (magenta dash-double-dotted).  
}%
\label{Fig:Press0}%
\end{figure}

In Fig.~\ref{Fig:Press0}  we show the contributions from the mesons ($\pi,{K},a_0,\kappa$) and partons ($u,d,s$ quarks and gluons) to the total pressure in hot quark matter as functions temperature for different values of the ratio $\mu/T$.
 We show the pressure in units of the Stefan-Boltzmann limit for the quark-gluon plasma ($N_c=N_f=3$)
 \begin{eqnarray}
 P_{\rm SB} &=& \frac{\pi^2}{90}T^4\left\{2(N_c^2-1)\right.\nonumber\\
 &&\left.+N_c\sum_{f=u,d,s}\left[ \frac{7}{2} +\frac{15}{\pi^2}\left(\frac{\mu_f}{T}\right)^2 + \frac{15}{2\pi^4}\left(\frac{\mu_f}{T}\right)^4\right]\right\}\ .\nonumber\\
 \label{P_SB}
 \end{eqnarray}
At finite temperature and  vanishing $\mu/T$ the partial pressure of the mesons shows a typical behavior, 
first increasing with temperature, then, when the chiral phase transition occurs, 
decreasing even before the Mott temperature is reached.
Above the Mott temperature, the growing meson width leads to a stronger reduction of the pressure with a rather sharp onset of this effect.  
In case of finite chemical potentials, along trajectories with fixed ratio $\mu/T$, the total pressure grows faster with increasing temperature because the chiral transition temperature drops with increasing $\mu/T$. 

\subsection{A possible explanation of the "horn effect" for the $K^+/\pi^+$ ratio?}
We want to come back to the observation made in the discussion of the phase shifts for the kaons 
shown in Fig.~\ref{Fig:ShiftsKaKaV} and their scalar partner states shown in Fig.~\ref{Fig:ShiftsKapKV1} 
that at finite baryon density and sufficiently high temperature a low-energy resonance or even bound state occurs in the positively charged channels and that is practically absent in the negatively charged channels.
The appearance of this anomalous mode is a consequence of the unequal masses of the constituents of these composite states and it is absent for the pions and their scalar partner states.
We are curious to see whether these anomalous states have a potential to contribute to the resolution of the puzzling observation of an enhancement of the ratio $K^+/\pi^+$ over the ratio $K^-/\pi^-$ of particle yields in heavy-ion collisions at $\sqrt{s_{NN}}\sim 8$ GeV (equivalent to $E_{\rm lab}\sim 30$ AGeV in fixed target experiments (the "horn" effect \cite{Gazdzicki:1998vd}), see \cite{Cleymans:2004hj} for the references to the experimental data and an early attempt to explain the location of the "horn" within a statistical model.
This work suggests that the transition from baryon to meson dominated entropy density may explain the position of the peak for the $K^+/\pi^+$ ratio. The sharpness of the peak, however, is not well reproduced by the statistical model. 
While standard kinetic approaches to particle production in heavy-ion collisions have failed to explain the horn effect, the inclusion of chiral symmetry restoration effects for the string decay in the PHSD approach recently resulted in a striking improvement \cite{Palmese:2016rtq}.
Chiral symmetry restoration as the precondition for the Mott dissociation of hadrons has also been the key element in an alternative attempt to provide a mechanism for chemical freeze-out 
\cite{Blaschke:2011ry,Blaschke:2011hm} and for the "horn" effect by Mott-Anderson localization \cite{Naskret:2015pna}.

\begin{figure}[!th]
\includegraphics[width=0.5\textwidth]{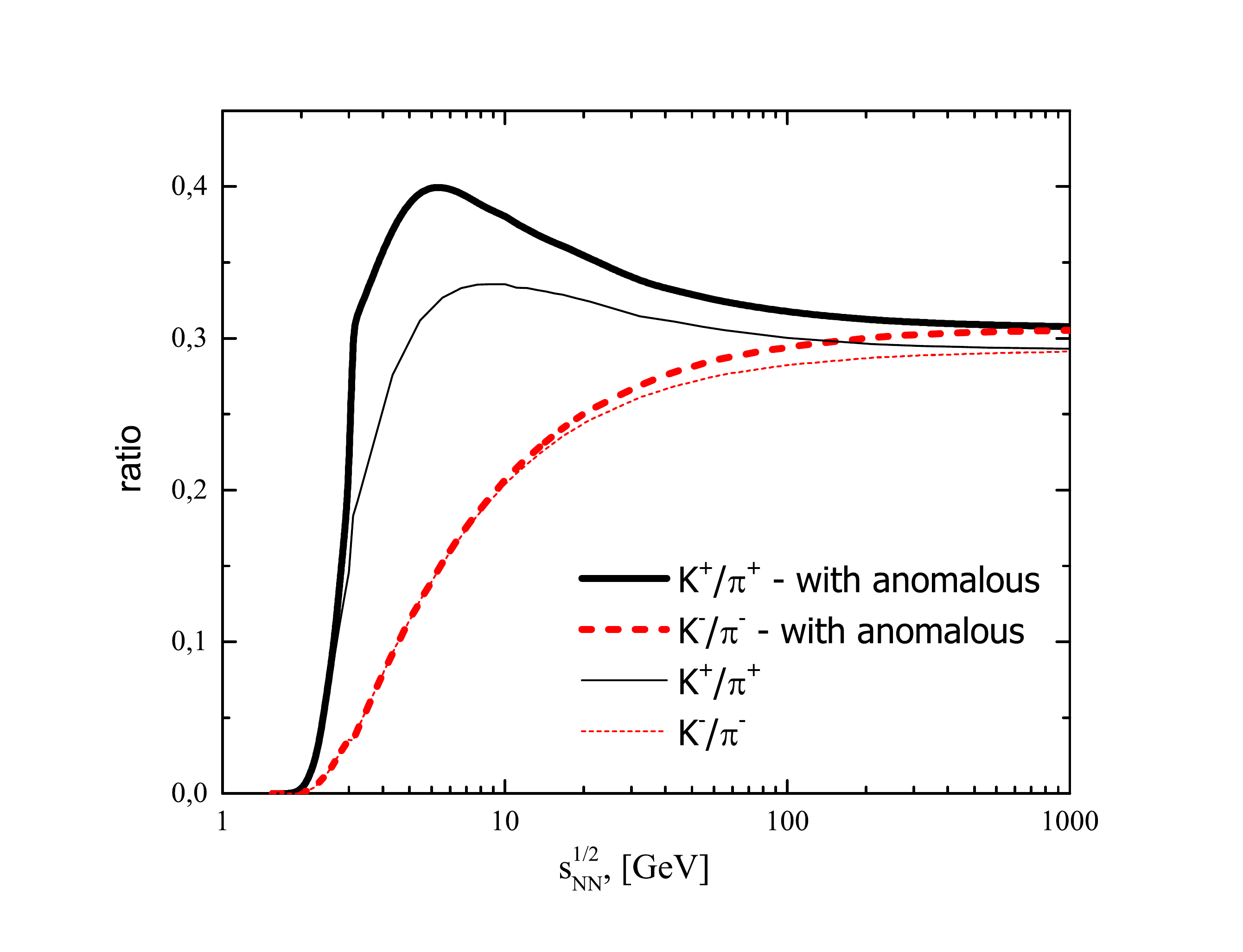}
\caption{
Ratio of yields $K^+/\pi^+$ and $K^-/\pi^-$ with and without the anomalous low-energy states as a function of the nucleon-nucleon center of mass energy $\sqrt{s_{NN}}$ in heavy-ion collisions.
}%
\label{Fig:K+pi+}%
\end{figure}

On the basis of the present approach to in-medium phase shifts for quark-antiquark scattering in hot, dense quark matter and the resulting mesonic contributions to the thermodynamics described within the Beth-Uhlenbeck approach, we propose to consider the ratios of partial pressures for the meson states according to Eq.~(\ref{BUU}) for describing the ratio of meson yields at freeze-out
\begin{eqnarray}
\frac{n_{K^\pm}}{n_{\pi^\pm}} = 
\frac{\int dM \int d^3p\  (M/E)[{\rm e}^{(E\mp \mu_K)/T}-1]^{-1}\delta_{K^\pm}(M)}
{\int dM \int d^3p\ (M/E)[{\rm e}^{(E - \mu_\pi)/T}-1]^{-1}\delta_{\pi^\pm}(M)}\ , 
\nonumber
\end{eqnarray}
where temperature and baryochemical potential are related to the collision energy $\sqrt{s_{NN}}$ by the
fit formula for the freeze-out parameters in the statistical model given, e.g., by \cite{Cleymans:2005xv}.
The nonequilibrium pion chemical potential is chosen as $\mu_pi=120$ MeV \cite{Kataja:1990tp}.
Since in the present model the absolute value of the pseudocritical temperature is larger than in lattice QCD, we propose to rescale the freeze-out temperature by the corresponding factor and to keep the ratio 
$\mu/T$. With this prescription we obtain the energy scan for the ratios $K^+/\pi^+$ and $K^-/\pi^-$
shown in Fig.~\ref{Fig:K+pi+}.
In order to highlight the possible role of the anomalous low-energy states for explaining the "horn" effect 
we show by thin lines the same ratios without these states. 

\section{Conclusions}
\label{sec:conclusion}

In this work we have described the thermodynamics of $N_f=2+1$ meson-quark-gluon matter at finite 
temperature and chemical potential in the framework of the Beth-Uhlenbeck approach.
In this formulation the quark-antiquark correlations in scalar and pseudoscalar channels are accounted for by the corresponding  phase shifts as solutions of the Bethe-Salpeter equations for the meson propagators. The notion of the phase shifts is generalized such as to display both the bound and scattering states spectrum with their characteristic medium effects. 
Most prominently this concerns the Mott transition where the bound state transforms to a resonance in the continuum. 
This transition manifests itself in the vanishing of the binding energy as a consequence of the lowering of the continuum edge of scattering states due to chiral symmetry restoration. 
It appears as a jump of the phase shift at threshold by $\pi$ in accordance with the Levinson theorem. 
Thus the mesonic pressure expressed in the BU form reflects this Mott dissociation effect. 
While the quark and gluon contribution to the total pressure is increased towards the Stefan-Boltzmann limit above the chiral restoration, the mesonic contribution to the pressure dies out. 

In the behaviour of the phase shifts for $K^\pm$ and $\kappa^\pm$ mesons we obtain an anomalous low-energy mode that is particularly pronounced for the positive charge states at finite densities and temperatures. 
We have discussed this phenomenon as a possible explanation for the "horn" effect, a pronounced peak in the energy scan of the $K^+/\pi^+$ ratio in heavy-ion collisions around $\sqrt{s_{NN}}\sim 8$ GeV. 
This behavior of the pressure in the meson-quark-gluon system at finite temperature and chemical potential is in qualitative agreement with the results from lattice QCD simulations, for the most recent ones see 
\cite{Borsanyi:2013bia,Bazavov:2014pvz}.
Quantitatively, there are differences the origin of which is well understood. 
The absolute value of the temperature for which the pressure starts rising towards the Stefan-Boltzmann limit is well above the one obtained in lattice QCD.
This discrepancy can be traced to the difference in the pseudocritical temperatures for chiral symmetry restoration. 
Local NJL and PNJL models predict chiral restoration temperatures above 200 MeV, while lattice QCD simulations give $T_c=154\pm 9 $ MeV \cite{Bazavov:2011nk}.
The so-called "entanglement" PNJL model has been suggested to cure this problem \cite{Sakai:2010rp}
by modifying the scalar coupling by a function of the traced Polyakov loop $G_S \to G_S(\Phi,\bar{\Phi})$.
Another improvement of the PNJL model concerns its nonlocal generalization which not only provides a solution to the problem of the pseudocritical temperature absolute value
\cite{Radzhabov:2010dd,Horvatic:2010md},
but also describes the running of the quark mass and the wave function renormalization of the quark propagator, in very close agreement with lattice QCD simulations at $T=0$ \cite{Bowman:2005vx}. 
At high temperatures, lattice QCD thermodynamics does not approach the massless Stefan-Boltzmann limit but rather a modified one which is well decribed by thermal masses in the two-loop approximation to the thermodynamic potential, see  \cite{Blaizot:1999ap} and references therein.
In the transition region, it is a matter of a detailed quantitative comparison, how many more hadronic resonances might be of relevance for a satisfactory description of the lattice QCD data.
It is clear that for very low temperatures and finite baryon densities the inclusion of baryonic states is customary. 
Then the EoS constraints from compact star phenomenology and heavy-ion collisions \cite{Klahn:2006ir}
shall be the guideline for the model development where lattice QCD data are absent.
These issues will be subject of subsequent work along the lines of the Beth-Uhlenbeck approach to the thermodynamics of the hadron-to-quark-gluon matter transition.
 
\section*{Acknowledgements}		
 The work of A.D. was supported by the Polish National Science Centre  (NCN) under grant number UMO - 2014/13/B/ST9/02621 and the Institute for Theoretical Physics of the University of Wroclaw under contract No. 1439/M/IFT/15. 
 D.B. and A.R.  are grateful for support from the NCN under grant number UMO-2011/02/A/ST2/00306.



\begin{thebibliography}{99}


\bibitem{Hufner:1994ma} 
  J.~H\"ufner, S.~P.~Klevansky, P.~Zhuang and H.~Voss,
  Annals Phys.\  {\bf 234}, 225 (1994).

\bibitem{Zhuang:1994dw} 
  P.~Zhuang, J.~H\"ufner and S.~P.~Klevansky,
  Nucl.\ Phys.\ A {\bf 576}, 525 (1994).

\bibitem{Blaschke:2013zaa} 
  D.~Blaschke,  M.~Buballa, A.~Dubinin, G.~R\"opke and D.~Zablocki,
Annals Phys.\  {\bf 348}, 228 (2014).

\bibitem{Dubinin:2013yga}
  A.~Dubinin, D.~Blaschke and Y.~L.~Kalinovsky,
  Acta Phys.\ Polon.\ Supp.\  {\bf 7} 1,  215 (2014).

\bibitem{Wergieluk:2012gd}
  A.~Wergieluk, D.~Blaschke, Y.~L.~Kalinovsky and A.~Friesen,
  Phys.\ Part.\ Nucl.\ Lett.\  {\bf 10}, 660 (2013).

\bibitem{Yamazaki:2012ux} 
  K.~Yamazaki and T.~Matsui,
  Nucl.\ Phys.\ A {\bf 913}, 19 (2013).

\bibitem{Yamazaki:2013yua} 
  K.~Yamazaki and T.~Matsui,
  Nucl.\ Phys.\ A {\bf 922}, 237 (2014).


\bibitem{Blaschke:2014zsa}
  D.~Blaschke, A.~Dubinin and M.~Buballa,
  Phys.\ Rev.\ D {\bf 91} 12,  125040 (2015).

\bibitem{Dubinin:2015glr} 
  A.~Dubinin, D.~Blaschke and A.~Radzhabov,
  J.\ Phys.\ Conf.\ Ser.\  {\bf 668}, no. 1, 012052 (2016).

\bibitem{Torres-Rincon:2016ahl} 
  J.~M.~Torres-Rincon and J.~Aichelin,
  arXiv:1601.01706 [nucl-th].

\bibitem{Ratti:2011au}
  C.~Ratti, R.~Bellwied, M.~Cristoforetti and M.~Barbaro,
  Phys.\ Rev.\ D {\bf 85}, 014004 (2012).

\bibitem{Nahrgang:2013xaa} 
  M.~Nahrgang, J.~Aichelin, P.~B.~Gossiaux and K.~Werner,
  Phys.\ Rev.\ C {\bf 89}, 014905 (2014). 

\bibitem{Nahrgang:2016lst} 
  M.~Nahrgang, J.~Aichelin, P.~B.~Gossiaux and K.~Werner,
  Phys.\ Rev.\ C {\bf 93}, no. 4, 044909 (2016).

\bibitem{Blaschke:2003ut} 
  D.~B.~Blaschke and K.~A.~Bugaev,
  Fizika B {\bf 13}, 491 (2004).

\bibitem{Turko:2011gw} 
  L.~Turko, D.~Blaschke, D.~Prorok and J.~Berdermann,
  Acta Phys.\ Polon.\ Supp.\  {\bf 5}, 485 (2012).

\bibitem{Blaschke:2015nma} 
  D.~Blaschke, A.~Dubinin and L.~Turko,
  Phys.\ Part.\ Nucl.\  {\bf 46}, no. 5, 732 (2015).


\bibitem{Dashen:1969ep} 
  R.~Dashen, S.~-K.~Ma and H.~J.~Bernstein,
  Phys.\ Rev.\  {\bf 187}, 345 (1969).

\bibitem{Costa:2002gk} 
  P.~Costa, M.~C.~Ruivo and Y.~L.~Kalinovsky,
  Phys.\ Lett.\ B {\bf 560}, 171 (2003).

\bibitem{Costa:2003uu}
  P.~Costa, M.~C.~Ruivo, Yu.~L.~Kalinovsky and C.~A.~de Sousa,
  Phys.\ Rev.\  C {\bf 70}, 025204 (2004).

\bibitem{Costa:2005cz}
  P.~Costa, M.~C.~Ruivo, C.~A.~de Sousa and Yu.~L.~Kalinovsky,
  Phys. Rev. {\bf D70}, 116013  (2004);
Phys.\ Rev.\  D {\bf 71}, 116002 (2005).


\bibitem{Rehberg:1995kh}
  P.~Rehberg, S.~P.~Klevansky and J.~H\"ufner,
  Phys.\ Rev.\  C {\bf 53}, 410 (1996).

\bibitem{Hansen:2006ee}
  H.~Hansen, W.~M.~Alberico, A.~Beraudo, A.~Molinari, M.~Nardi and C.~Ratti,
  Phys.\ Rev.\  D {\bf 75}, 065004 (2007).

\bibitem{Costa:2008dp}
  P.~Costa, M.~C.~Ruivo, C.~A.~de Sousa, H.~Hansen and W.~M.~Alberico,
  Phys.\ Rev.\  D {\bf 79}, 116003 (2009).

\bibitem{Ratti:2005jh}
  C.~Ratti, M.~A.~Thaler and W.~Weise,
  Phys.\ Rev.\  D {\bf 73}, 014019 (2006).

\bibitem{Roessner:2006xn}
  S.~Roessner, C.~Ratti and W.~Weise,
  Phys.\ Rev.\  D {\bf 75}, 034007 (2007).

\bibitem{Klevansky:1992qe} 
  S.~P.~Klevansky,
  Rev.\ Mod.\ Phys.\  {\bf 64}, 649 (1992).

\bibitem{Naskret:2015pna}
  M.~Naskret, D.~Blaschke and A.~Dubinin,
  Phys.\ Part.\ Nucl.\  {\bf 46}, no. 5,  789 (2015).
  
\bibitem{Karsch:2010ck} 
  F.~Karsch and K.~Redlich,
  Phys.\ Lett.\ B {\bf 695}, 136 (2011).
  
\bibitem{Gazdzicki:1998vd} 
  M.~Gazdzicki and M.~I.~Gorenstein,
  Acta Phys.\ Polon.\ B {\bf 30}, 2705 (1999).

\bibitem{Cleymans:2004hj} 
  J.~Cleymans {\it et al.} [NA49 Collaboration],
  Phys.\ Lett.\ B {\bf 615}, 50 (2005).
  
\bibitem{Palmese:2016rtq} 
  A.~Palmese, W.~Cassing, E.~Seifert, T.~Steinert, P.~Moreau and E.~L.~Bratkovskaya,
  arXiv:1607.04073 [nucl-th].
  
\bibitem{Blaschke:2011ry} 
  D.~B.~Blaschke, J.~Berdermann, J.~Cleymans and K.~Redlich,
  Phys.\ Part.\ Nucl.\ Lett.\  {\bf 8}, 811 (2011).
  
\bibitem{Blaschke:2011hm} 
  D.~Blaschke, J.~Berdermann, J.~Cleymans and K.~Redlich,
  Few Body Syst.\  {\bf 53}, 99 (2012).

\bibitem{Cleymans:2005xv} 
  J.~Cleymans, H.~Oeschler, K.~Redlich and S.~Wheaton,
  Phys.\ Rev.\ C {\bf 73}, 034905 (2006).

\bibitem{Kataja:1990tp} 
  M.~Kataja and P.~V.~Ruuskanen,
  Phys.\ Lett.\ B {\bf 243}, 181 (1990).
  
\bibitem{Borsanyi:2013bia}
  S.~Borsanyi, Z.~Fodor, C.~Hoelbling, S.~D.~Katz, S.~Krieg and K.~K.~Szabo,
  Phys.\ Lett.\ B {\bf 730}, 99 (2014).

\bibitem{Bazavov:2014pvz}
  A.~Bazavov {\it et al.} [HotQCD Collaboration],
  Phys.\ Rev.\ D {\bf 90}, no. 9, 094503 (2014).
    
\bibitem{Bazavov:2011nk} 
  A.~Bazavov {\it et al.},
  Phys.\ Rev.\ D {\bf 85}, 054503 (2012).
 
\bibitem{Sakai:2010rp} 
  Y.~Sakai, T.~Sasaki, H.~Kouno and M.~Yahiro,
  Phys.\ Rev.\ D {\bf 82}, 076003 (2010).
    
\bibitem{Radzhabov:2010dd} 
  A.~E.~Radzhabov, D.~Blaschke, M.~Buballa and M.~K.~Volkov,
  Phys.\ Rev.\ D {\bf 83}, 116004 (2011).
    
\bibitem{Horvatic:2010md} 
  D.~Horvatic, D.~Blaschke, D.~Klabucar and O.~Kaczmarek,
  Phys.\ Rev.\ D {\bf 84}, 016005 (2011).
        
\bibitem{Bowman:2005vx} 
  P.~O.~Bowman et al., 
  Phys.\ Rev.\ D {\bf 71}, 054507 (2005).
    
\bibitem{Blaizot:1999ap} 
  J.~P.~Blaizot, E.~Iancu and A.~Rebhan,
  Phys.\ Lett.\ B {\bf 470}, 181 (1999).
    
\bibitem{Klahn:2006ir} 
  T.~Kl\"ahn {\it et al.},
  Phys.\ Rev.\ C {\bf 74}, 035802 (2006).
    
\end{thebibliography}
\end{document}